%% file: main.tex
\documentclass[USenglish,online]{article}

\ifx\directlua\undefined\ifx\XeTeXcharclass\undefined
  \else\RequirePackage[no-math]{fontspec}[2017/03/31]\fi 
  \else\RequirePackage[no-math]{fontspec}[2017/03/31]\fi 
\usepackage[big]{dgruyter}

\usepackage{wrapfig}
\usepackage{xcolor}
\usepackage{booktabs}
\usepackage{comment}
\usepackage{graphicx}
\usepackage{amssymb}
\usepackage{amsmath}
\usepackage{pifont}
\usepackage{hyperref}

\newcolumntype{L}[1]{>{\raggedright\let\newline\\\arraybackslash\hspace{0pt}}m{#1}}
\newcolumntype{C}[1]{>{\centering\let\newline\\\arraybackslash\hspace{0pt}}m{#1}}
\newcolumntype{R}[1]{>{\raggedleft\let\newline\\\arraybackslash\hspace{0pt}}m{#1}}

\newcommand{\cmark}{\ding{51}}%
\newcommand{\xmark}{\ding{55}}%

\newcommand{\JPO}[1]{\textcolor{teal}{JPO: #1}}

\newcommand{\EB}[1]{\textcolor{orange}{EB: #1}}
\newcommand{\AH}[1]{\textcolor{blue}{AH: #1}}

\begin{document}

\articletype{Research Article}
\author*[1]{Christian Neurohr}
\author[2]{Marcel Saager}
\author[2]{Lina Putze}
\author[2]{Jan-Patrick Osterloh}
\author[2]{Karina Rothemann}
\author[2]{Hilko Wiards}
\author[2]{Eckard Böde} 
\author[2]{Axel Hahn}
\runningauthor{Neurohr et al.}
\affil[1]{German Aerospace Center (DLR e.V.), Institute of Systems Engineering for Future Mobility, Oldenburg, Germany, E-mail: christian.neurohr@dlr.de}
\affil[2]{German Aerospace Center (DLR e.V.), Institute of Systems Engineering for Future Mobility, Oldenburg, Germany, E-mail: vorname.nachname@dlr.de}
\title{Towards Efficient Certification of Maritime Remote Operation Centers -- Ansatz zur effizienten Zertifizierung von maritimen Fernsteuerungszentren}
\runningtitle{Towards Efficient Certification of Maritime ROCs}
\abstract{Additional automation being build into ships implies a shift of crew from ship to shore. However, automated ships still have to be monitored and, in some situations, controlled remotely. These tasks are carried out by human operators located in shore-based remote operation centers.
In this work, we present a concept for a hazard database that supports the safeguarding and certification of such remote operation centers. The concept is based on a categorization of hazard sources which we derive from a generic functional architecture. A subsequent preliminary suitability analysis unveils which methods for hazard analysis and risk assessment can adequately fill this hazard database. -- \\ Die zunehmende Automatisierung von Schiffen führt zu einer Verlagerung der Besatzung vom Schiff ans Land. Automatisierte Schiffe müssen jedoch weiterhin überwacht und in bestimmten Situationen ferngesteuert werden. Diese Aufgaben werden von menschlichen Operateuren in landgestützten Fernsteuerungszentren ausgeführt.
In dieser Arbeit stellen wir ein Konzept für eine Gefährdungsdatenbank vor, welche die Sicherung und Zertifizierung solcher Fernsteuerungszentren unterstützt. Das Konzept basiert auf einer Kategorisierung von Gefährdungsquellen, abgeleitet aus einer generischen Funktionsarchitektur. Eine anschließende vorläufige Eignungsanalyse zeigt, welche Methoden zur Gefährdungsanalyse und Risikobewertung diese Gefährdungsdatenbank füllen können.}

\keywords{Remote Operation Center, Autonomous Surface Ships, Hazard Analysis \& Risk Assessment, Safety, Certification, Human Factors -- Fernsteuerungszentrum, Autonome Überwasserschiffe, Gefährdungsanalyse und Risikobewertung, Sicherheit, Zertifizierung, Menschfaktoren}
\received{...}
\accepted{...}
\journalname{at - Automatisierungstechnik}
\journalyear{...}
\journalvolume{..}
\journalissue{..}
\startpage{1}
\aop
\DOI{...}

\maketitle

\section{Introduction}
\label{sec:introduction}

\input{sections/01_Introduction.tex} 

\section{Related Work}
\label{sec:related_work}
\input{sections/02_Related_work.tex} 

\section{Maritime Remote Operation Centers}
\label{sec:remote_operation_centers}
\input{sections/03_ROC}

\section{Building a Hazard Database for Remote Operation Centers}
\label{sec:hazard_database}
\input{sections/04_HARA}

\section{Conclusion}
\label{sec:conclusion}
\input{sections/10_Conclusion.tex}

\newpage
\bibliographystyle{plain}
\bibliography{Literature.bib}

\newpage
\section*{Authors' Biographies}

\begin{wrapfigure}{l}{25mm} 
    \includegraphics[width=1in,height=1.25in,clip,keepaspectratio]{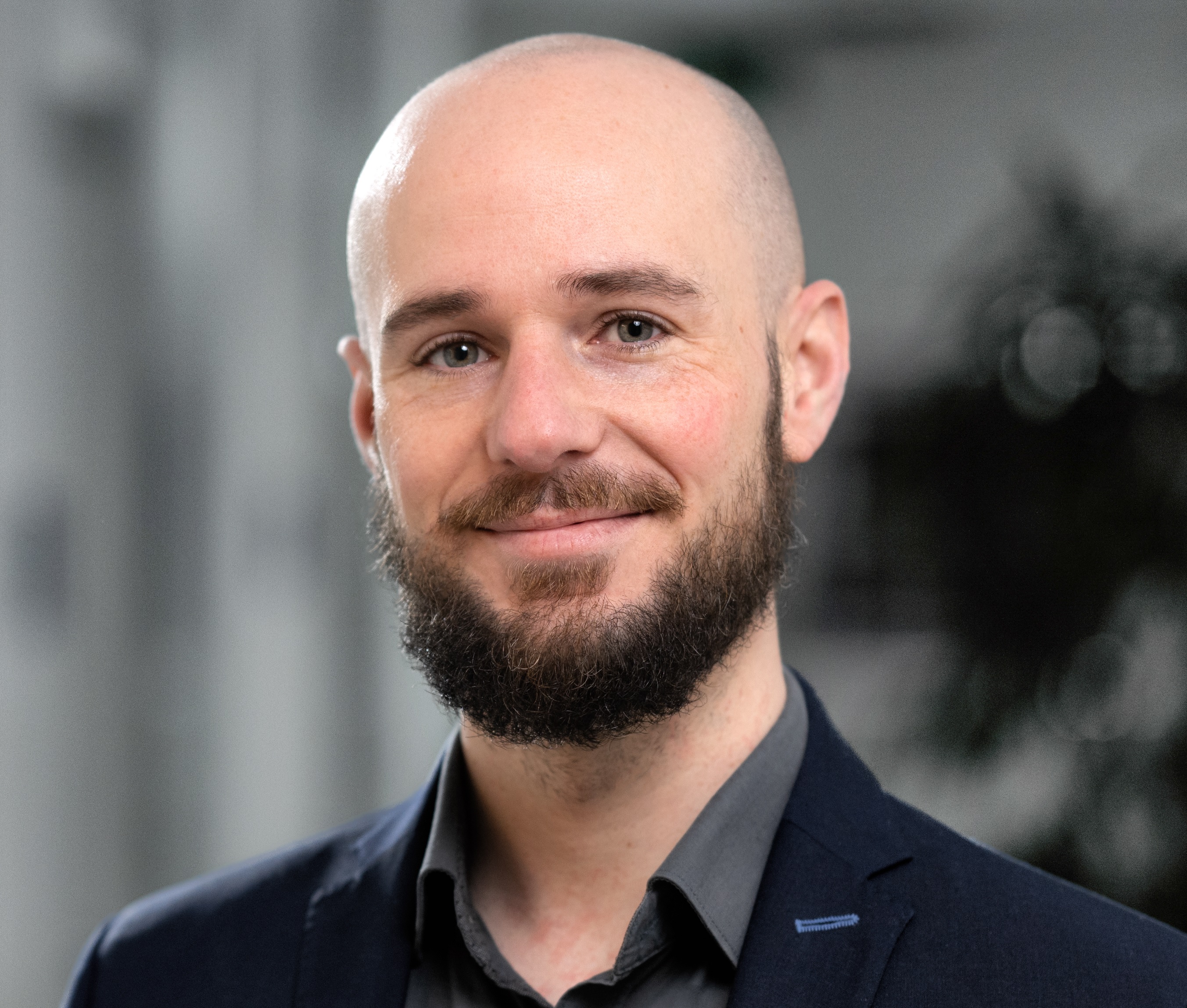}
 \end{wrapfigure}\par
\textbf{Christian Neurohr} received the B.Sc.\ and M.Sc.\ in mathematics in 2011 and 2013 from RPTU Kaiserslautern, Germany and his Ph.D.\ (Dr. rer. nat.) from Carl von Ossietzky Universit\"at Oldenburg, Germany in 2018.
After a short period as a visiting researcher at the University of Sydney, he started his occupation as a postdoctoral researcher at the German Aerospace Center (DLR e.V.) Institute of Systems Engineering for Future Mobility where he is working in the area of scenario-based verification and validation of automated vehicles. Since 2023 he leads the 'Criticality Analysis' team within the division 'Theory and Design'.\par

\begin{wrapfigure}{l}{25mm} 
    \includegraphics[width=1in,height=1.25in,clip,keepaspectratio]{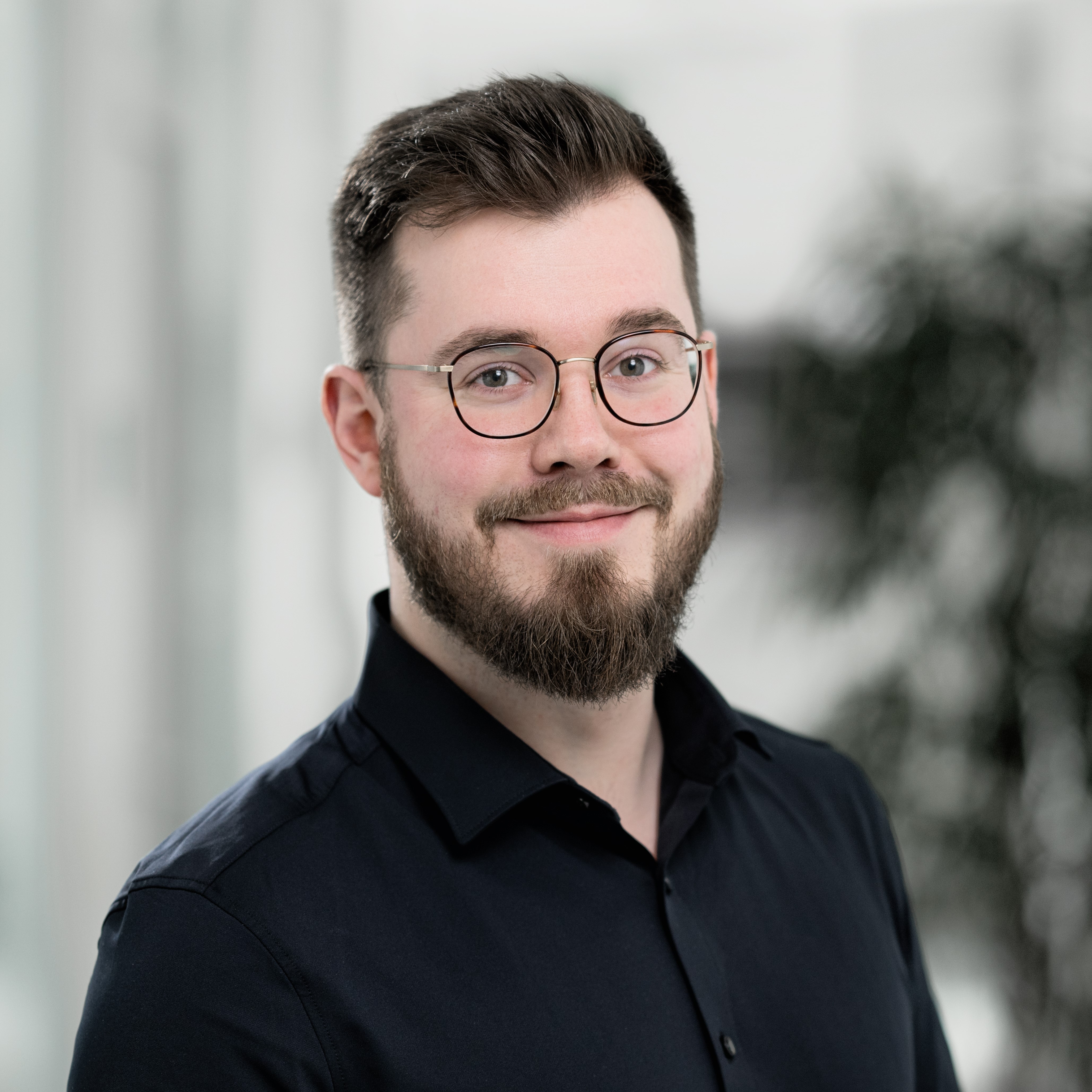}
 \end{wrapfigure}\par
\textbf{Marcel Saager} received the B.A.\ in Business and Economics and M.Sc.\ in Business Information Systems in 2016 and 2019  from Carl von Ossietzky Universit\"at Oldenburg, Germany.
After two years as a Modelling- and Software Engineer at Humatects, a company specialized in Human Machine Interaction Solutions, he started his occupation as a doctoral researcher at the German Aerospace Center (DLR e.V.) Institute of Systems Engineering for Future Mobility where he is working in the area of human factors and human centered engineering of highly automated vessels and trains. Furthermore he works and worked as a Lecturer at University of Oldenburg, Private University of Applied Sciences Vechta and University of Applied Sciences in Nuertingen-Geißlingen.\par

\begin{wrapfigure}{l}{25mm} 
    \includegraphics[width=1in,height=1.25in,clip,keepaspectratio]{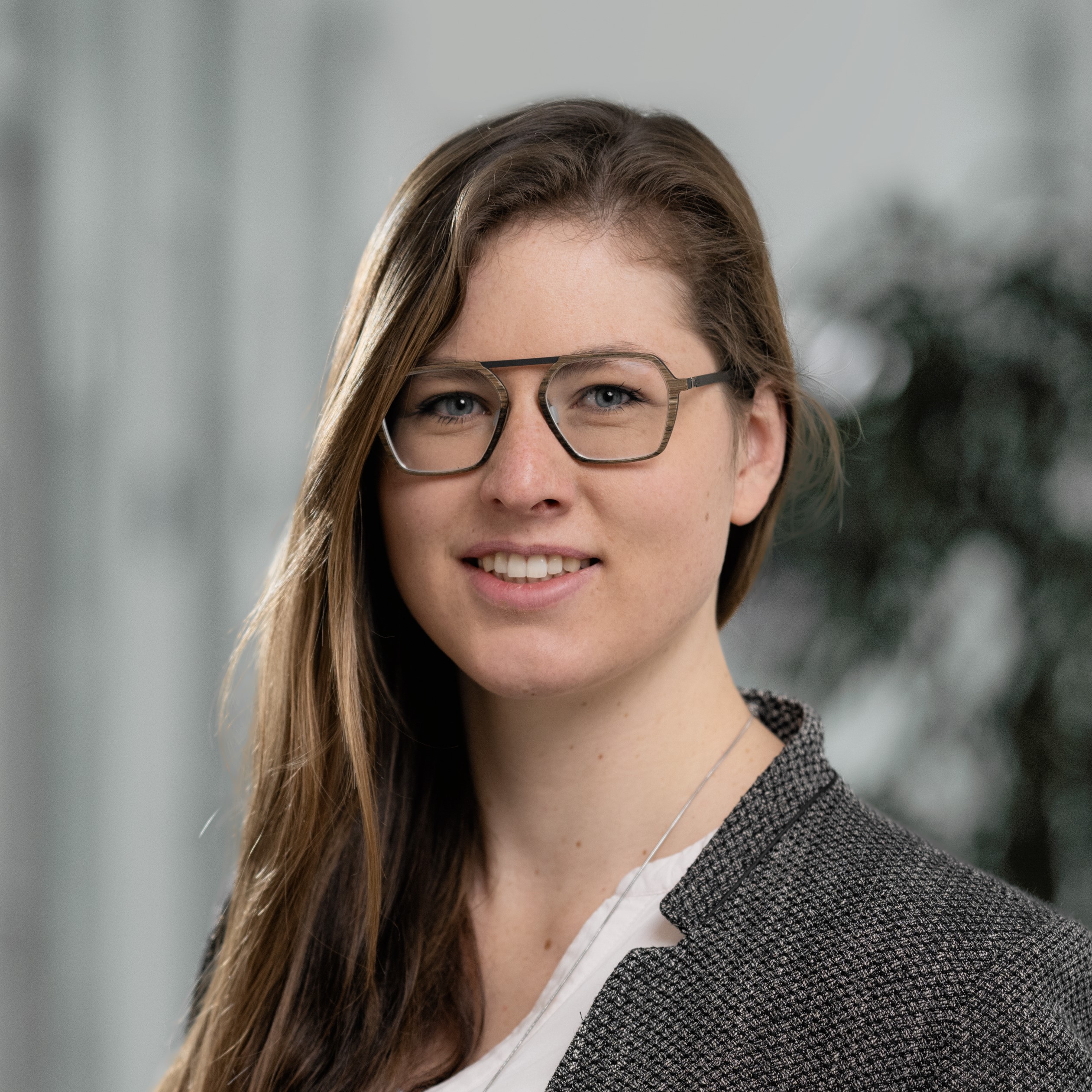}
 \end{wrapfigure}\par
\textbf{Lina Putze} received the B.Sc.\ and M.Sc.\ degrees in mathematics from
the University of Münster in 2016 and 2019, specializing on the topics
of stochastic processes, probability theory and its applications. She is
currently working as a researcher at the group System Concepts and Design Methods at the German Aerospace Center (DLR
e.V.) Institute of Systems Engineering for Future Mobility. The focus of
her research is on methods to ensure trustworthiness of highly automated transport
systems in different domains, including the identification and analysis of hazards and risk
triggering scenario properties, causal analysis and risk assessment. 

 \par

\begin{wrapfigure}{l}{25mm} 
    \includegraphics[width=1in,height=1.25in,clip,keepaspectratio]{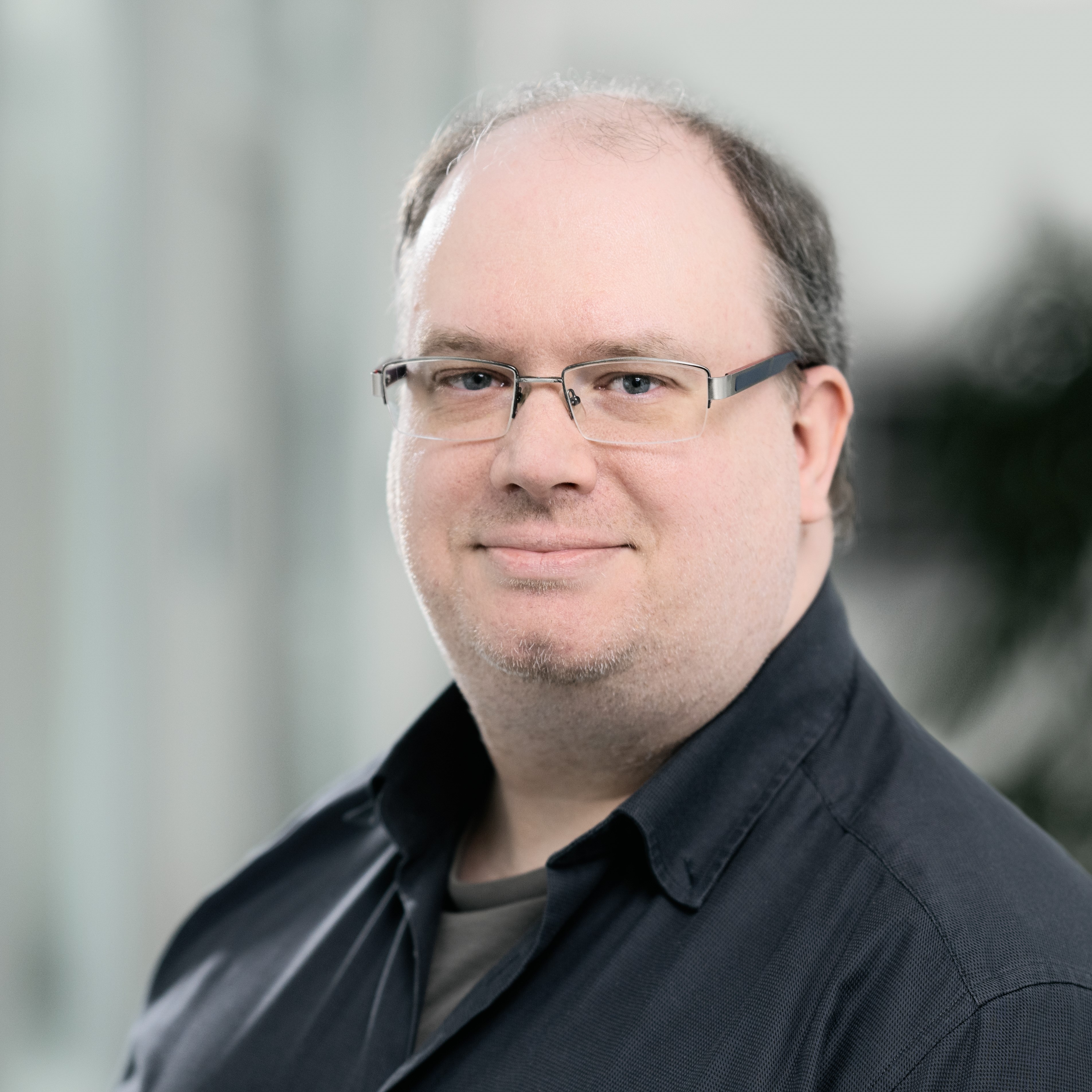}
\end{wrapfigure}\par
\textbf{Jan-Patrick Osterloh} received his Diploma in Computer Science in 2005 from the Carl von Ossietzky Universität Oldenburg, Germany, and began his professional career at the Human Centred Engineering Group within the Transportation Division of OFFIS. In the course of a structural reorganization, this division was transferred to the German Aerospace Center (DLR e.V.) and now forms the Institute of Systems Engineering for Future Mobility. As a Senior Research Engineer, his research focuses on human factors and cognitive modelling, particularly human error, perception, workload, and situation awareness in the aeronautics, automotive, and maritime domains. In addition to his work in Human Factors, he serves as the institute’s Software Engineering Contact, supporting both methodological and technical aspects of software development. 
\par

\begin{wrapfigure}{l}{25mm} 
    \includegraphics[width=1in,height=1.25in,clip,keepaspectratio]{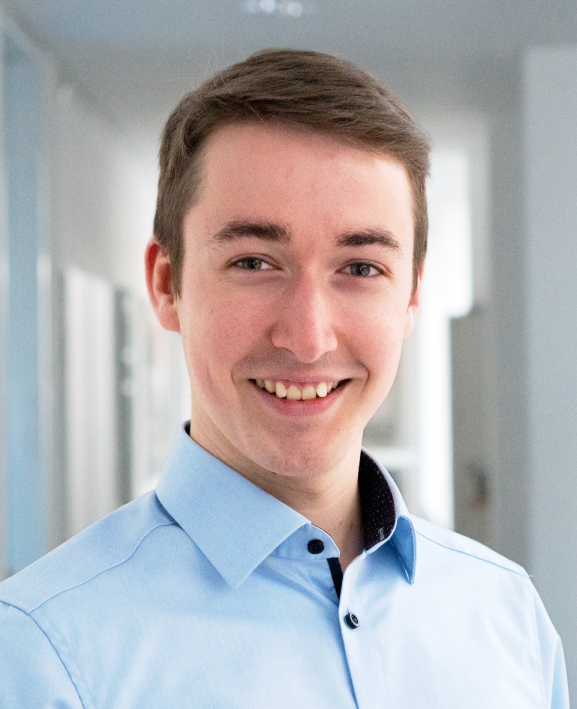}
 \end{wrapfigure}\par
\textbf{Hilko Wiards} received his B.Sc. and M.Sc. degrees in Computer Science in 2018 and 2020, respectively, from Carl von Ossietzky Universität Oldenburg, Germany. He began his professional career at the Cooperative Mobile Systems Group within the Transportation division at OFFIS - Institute for Information Technology. Following a structural reorganization, this division was transferred to the German Aerospace Center (DLR e.V.) and now constitutes the Institute of Systems Engineering for Future Mobility. Within the department of Safe Automation Maritime Systems his research focuses on the navigational aspects of autonomous and remotely operated vessels. This includes the evaluation of new sensor systems, fallback procedures and redundancies.

\par

\begin{wrapfigure}{l}{25mm} 
    \includegraphics[width=1in,height=1.25in,clip,keepaspectratio]{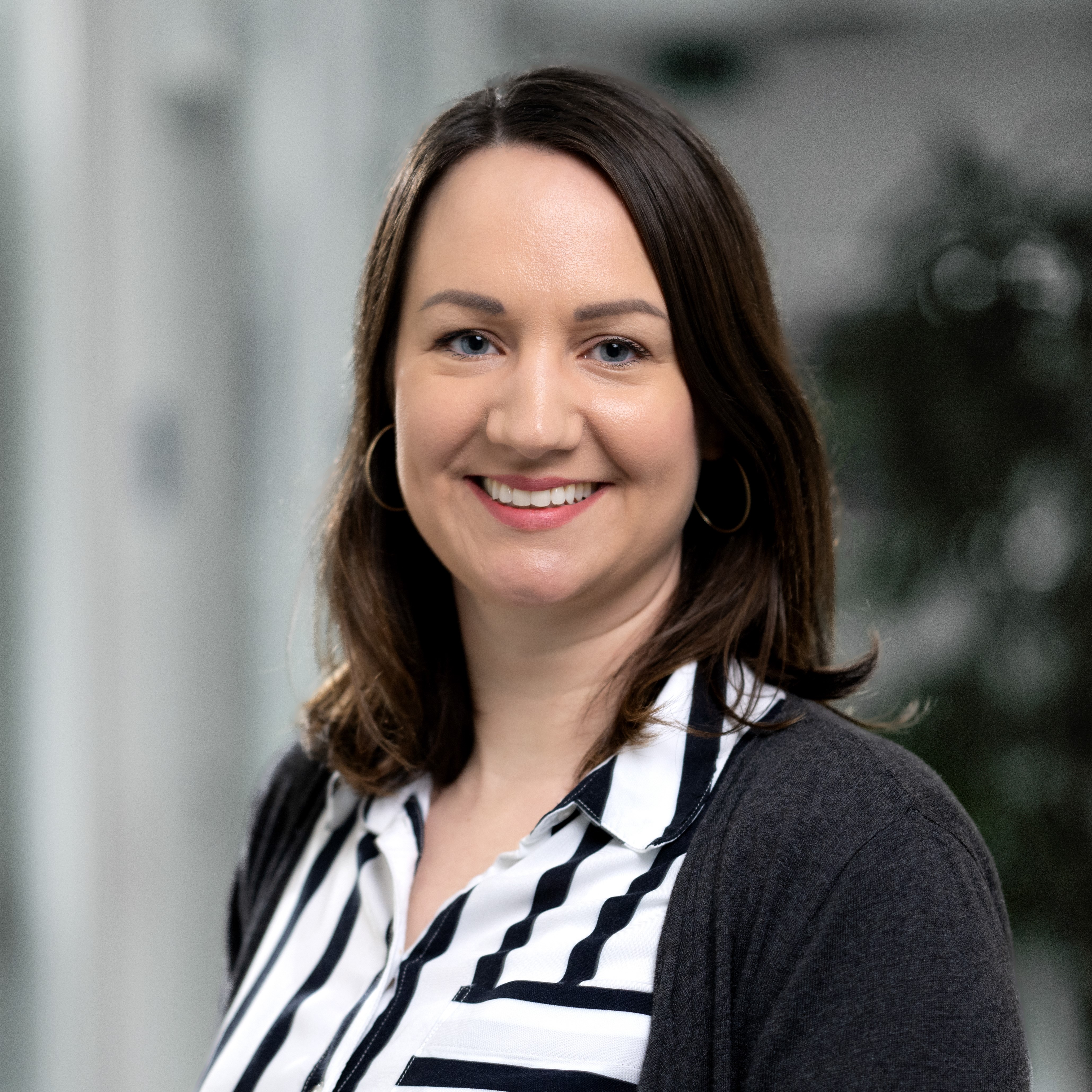}
 \end{wrapfigure}\par
\textbf{Karina Rothemann} received the B.Eng.\ in mechanical engineering and design and M.Eng.\ in mechanical engineering from Hochschule Emden/Leer in 2018 and 2020, specializing on the topcis of product and process optimization using AI components. She has worked in the research and development departments of several major automotive companies and starts her carrier in science at the OFFIS- Institut of Computer Science.Currently she works as a researcher at the German Aerospace Center (DLR e.V.) Institute of Systems Engineering for Future Mobility, where she is focusing on the application of hazard and risk analysis methods in the development process of highly automated systems.\par

\begin{wrapfigure}{l}{25mm} 
    \includegraphics[width=1in,height=1.25in,clip,keepaspectratio]{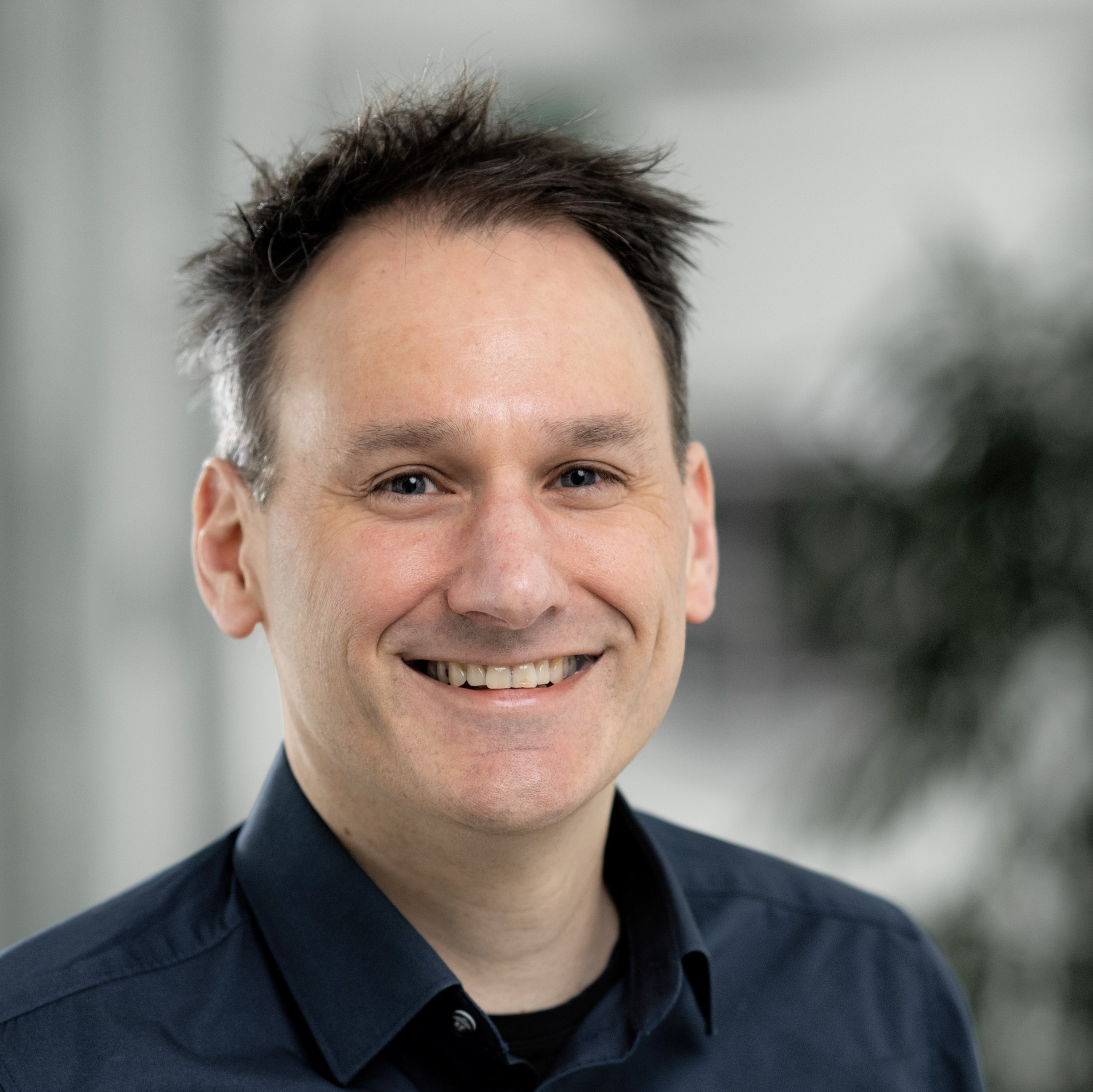}
 \end{wrapfigure}\par
\textbf{Eckard Böde} received his Dipl.-Inform.\ degree in Computer Science from the Carl von Ossietzky University, Oldenburg, Germany, in 2001. He subsequently joined OFFIS e.V., where he focused on safety assessment and model-based safety analysis for aerospace and automotive applications. In 2012, he was appointed Group Leader for Safety Analysis and Verification. He currently leads the R\&D group System Concepts and Design Methods at the German Aerospace Center (DLR e.V.), Institute of Systems Engineering for Future Mobility. His research interests include methods and tools for the design and verification of trustworthy cyber-physical systems, with a particular emphasis on safety assessment of automated systems and the integration of functional safety with SOTIF in safety cases.\par

\begin{wrapfigure}{l}{25mm} 
    \includegraphics[width=1in,height=1.25in,clip,keepaspectratio]{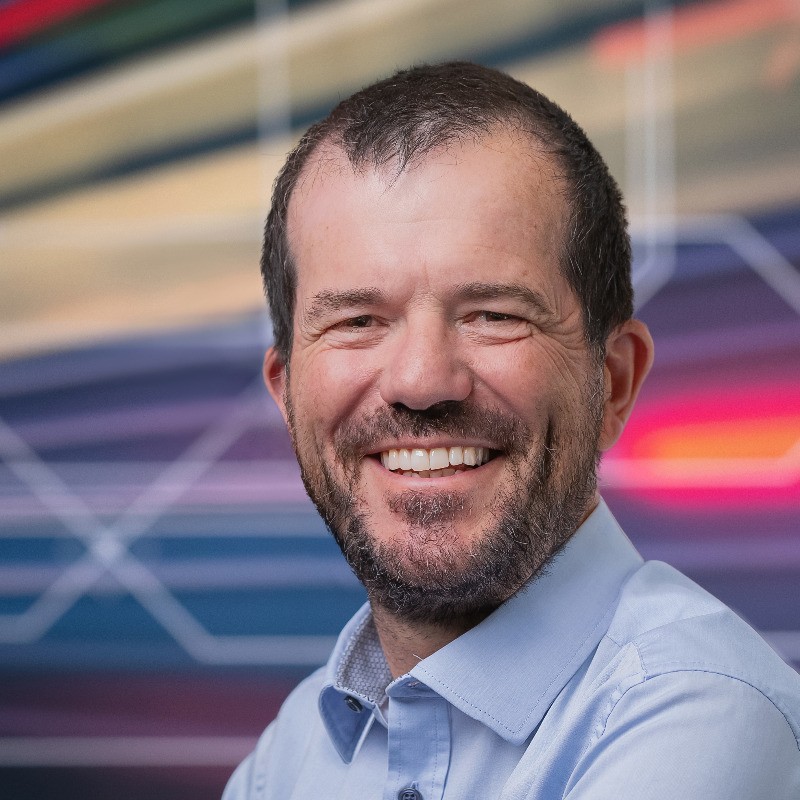}
 \end{wrapfigure}\par
\textbf{Axel Hahn} holds a Doctorate in Mechanical Engineering from the University of Paderborn. He currently serves as the Director of the German Aerospace Center (DLR e.V.) Institute of Systems Engineering for Future Mobility, which emerged from the former Transportation Division of OFFIS. In addition, he has a professorship at the Carl von Ossietzky University Oldenburg. His work centers on dependable and intelligent systems in mobility and transport, with a focus on software engineering, system architecture, and safety-critical applications across the automotive, maritime, and aeronautics domains. With a strong interdisciplinary orientation, he bridges research and practical innovation in digitization, automation, and systems engineering. He has led numerous national and European research projects and actively contributes to shaping future mobility concepts through both technical leadership and strategic guidance.
\par

\end{document}

%% file: sections/01_Introduction.tex
Increasing levels of automation is being build into vessels, leading to the advent of Maritime Autonomous Surface Ships (MASS), as defined by the Internation Maritime Organization (IMO) \cite{ISOTS23860}.
Shore-based Remote Operation Centers (ROCs) are emerging as a supplemental technology alongside MASS to monitor or remote-control such ships in harbors, in coastal waters, or for inland waterways \cite{gali2024remote,bratic2019review}.
There exist various novel approaches adapted for the Hazard Analysis and Risk Assessment (HARA) of MASS to cope with the complexity that is introduced by automation technology \cite{wylie2024safety}.
For example, the Risk-based Assessment Tool commissioned by the European Maritime Safety Agengy (EMSA) \cite{emsa2024rbat}. 
Regarding classification of MASS, there exists class guidelines such as the DNV-CG-0264 on  \emph{Autonomous and Remotely Operated Ships}.
These safeguarding approaches are centered around MASS and possible remote control components are considered as MASS functionality. Therefore, the safety of ROCs is considered as part of the MASS HARA.
In this concept paper, we put forth the idea to collect safety-relevant artifacts for a generic ROC in a database to enable their reuse. Hence, to facilitate more efficient certification of MASS-ROC pairs, we tackle the research question
\begin{center}
 \emph{How can we build a hazard database that integrates technical as well as human hazards and facilitates the efficient certification of generic ROC-MASS pairs with well-defined interfaces?}
\end{center}
In this regard, we contribute
\begin{itemize}
    \item[$\bullet$] a review of standards, regulations, and HARA methods in the maritime domain in \autoref{sec:related_work},
    \item[$\bullet$] a generic ROC-MASS functional architecture as a starting point for a hazard database in \autoref{sec:remote_operation_centers},
    \item[$\bullet$] a categorization of hazard sources for ROCs, a suitability analysis for methods to adequately cover these categories, and potential benefits of a hazard database in \autoref{sec:hazard_database}.
\end{itemize}

%% file: sections/02_Related_Work.tex

In this section, with a distinct focus on ROCs, we briefly review relevant standards and regulations in the maritime domain in \autoref{subsec:standards_regulations} and relevant HARA methods in \autoref{subsec:hara_methods}.

\subsection{Relevant Standards and Regulations}
\label{subsec:standards_regulations}

\textbf{IMO MASS Code}: Within the IMO, the Maritime Safety Committee (MSC) developed the \emph{Interim Guidelines for MASS Trials} in 2019 \cite{IMO2019}.
These interim guidelines seek to assist authorities and stakeholders to conduct trials for MASS and related systems safely, securely, and under environment protection.
These guidelines were specifically formulated for experimental trials over limited periods. The MASS Code is on track to be finalized in its non-mandatory form in 2026 \cite{IMO2025}. The code uses a goal-based approach, focusing on performance standards rather than prescriptive rules. 

\textbf{DNV-RU-SHIP}(Pt.6,Ch.12): The classification society DNV updated their rules for classification document in 2024 to introduce the Autonomous and Remotely Operated Ships (AROS) class notation \cite{DNVRUPT6CH12}. AROS as a framework wants MASS to be at least as safe as conventional vessels. Therein, the term \emph{autoremote} refers to operations, tasks, functions, or systems that enhance decision support, remote control, or autonomy compared to traditional crewed ships. 
There are four AROS notations -- navigation, engineering, operations, safety -- each with one qualifier for operation mode and one describing the control location.

\textbf{DNV-CG-0264}: The Class Guideline DNV-CG-0264 \cite{dnv2018autonomous}, developed by DNV, aims to provide guidance for the safe implementation of novel technologies in the context of autoremote vessel functions. 
Further, it formulates a recommended work process to obtain the approval of novel concepts. 
Overall, the framework is designed to ensure that the implementation of innovative concepts and technologies meets or exceeds the safety standards of traditional vessel operations. It also outlines the specific class notations applicable to autonomous and remotely operated ships (AROS), cf.~\cite{DNVRUPT6CH12}.


\textbf{DNV-ST-0324}: The DNV-ST-0324 standard provides a set of required competences for humans operating ROCs, i.e., operators tasked with supporting, monitoring, or controlling MASS from a remote, shore-based location \cite{dnv2022competence}. The standard makes suggestions on the necessary skills and knowledge for human ROC-operators regarding communication, navigation, machinery, and cargo. As such the standard is highly relevant to the identification of hazards (and their causes) related to human factors -- due to ROC-operator potentially lacking basic skills or knowledge. The DNV-ST-0324 is complemented by the recommended practice DNV-RP-0323 regarding certification schemes for ROC-operators \cite{dnv2021certification}.

\textbf{ISO 23860}: The technical specification ISO/TS 23860 covers vocabulary for highly automated or autonomous ships \cite{ISOTS23860}. It introduces terms ranging from related to autonomous ship systems, e.g., \emph{autonomy} and \emph{control}. In this work, we will follow the ISO/TS 23860's definition for terms such as ROC or MASS.
Furthermore, the standard addresses the interrelations of terms for autonomous ships such as the relations between of ROC, MASS, and support services.

\textbf{CMOROC}: Based on the ISO/TS 23860, the EMSA study ‘CMOROC - Identification of Competences for MASS Operators in Remote Operation Centres’ analyses the requirements for future operators of MASS from ROC. It is based on three representative ship types (feeder, RoPax ferry, bulk carrier) and examines which tasks are required in a ROC and which competences are necessary for this. The study identifies different levels of automation and develops a structured model for the operation and distribution of roles within a ROC. A key result is a catalogue of skills based on the STCW (Standards of Training, Certification and Watchkeeping for Seafarers), which is used as the basis for training. Building on this, a basic and an advanced curriculum is being developed \cite{emsa2023CMOROC}.

\textbf{ISO 26262 \& ISO 21448}: The ISO~26262 \cite{iso26262} and ISO~21448 \cite{iso21448} are safety standards from the automotive domain that complement each other. While the ISO~26262 focuses on functional safety -- addressing risks arising from system failures, ISO~21448 is concerned with the safety of the intended functionality (SOTIF). SOTIF addresses risks that arise not from system failures, but from insufficiencies in the specification, performance limitations or the inability to detect or prevent reasonably foreseeable misuses. ISO~21448 provides a structured framework that offers guidance on managing SOTIF for road vehicles equipped with automated driving systems. The standard organizes the main SOTIF activities and defines high-level objectives to support a systematic development and validation process \cite{putze2023validation}. Although the ISO~21448 specifically targets driving automation, the concept of SOTIF is also highly relevant for automated systems in other domains.

\subsection{Relevant Methods for Hazard Analysis and Risk Assessment}
\label{subsec:hara_methods}

For conventional vessels, there exists a wide range of well-established HARA methods which are commonly applied in practice, e.g., Failure Mode and Effect Analysis (FMEA) or Fault Tree Analysis (FTA). 
To address the specific challenges posed by automated maritime systems, such as their reliance on sensor perception and the complexity of system interactions, adaptions of established methods as well as new approaches have been explored in recent research \cite{wylie2024safety, Thieme2018, Li2023}. 
However, only a limited number of studies explicitly consider ROCs. A literature review provided by Zhou et al.~\cite{Zhou2020}, which evaluates the suitability of commonly used HARA methods for automated maritime system, highlights this gap. Notably, the authors observe that among the evaluated studies, so far only System-Theoretic Process Analysis (STPA) was applied  with explicit consideration of the communication between vessel and ROC \cite{Wrobel2018a, Wrobel2018b, Banda2019, Utne2017}.
Similarly, another literature review on risk models for automated maritime systems by Thieme et al.~\cite{Thieme2018} reports that only two of the investigated studies explicitly address the communication with a ROC -- one employing STPA, and the other a combination of brainstorming and Bayesian Networks \cite{Wrobel2018a, Wrobel2016}. 
Furthermore, Li et al.\ emphasize the importance of incorporating human factors into the HARA of automated maritime systems, noting that these systems constitute highly complex socio-technical systems in which the role of the remote operator is significantly more complex than that of a traditional onboard operator \cite{Li2023}.

In the following, we briefly introduce common HARA methods that may be applicable for ROCs, as well as some emerging approaches specifically developed for highly automated systems.
As the human operator is of particular relevance for remote operation, we consider not only HARA methods focusing on technical system safety but also methods that explicitly address human factors.

\subsection*{Technical System Safety}

There exists a variety of methods for identifying and analyzing hazards arising from faults or insufficiencies within the system, each with a slightly different focus. Some approaches emphasize the identification of component failures, while others concentrate on the analysis of causal chains or the evaluation of the associated risks. The methods employ either inductive or deductive reasoning strategies and can be tailored for specific stages of the system development process. Moreover, HARA techniques can be roughly classified as qualitative or quantitative, depending on whether they primarily rely on expert judgment or derive probabilistic statements from data.

A broadly applied hazard analysis method is the \textbf{Failure mode and effects analysis (FMEA)} \cite{SAEJ1739}.
FMEA follows a seven-stage procedure, that focuses on identifying failure modes, their causes and effects on the overall system. The method relies on inductive reasoning and emphasizes systematic documentation throughout the analysis process. Typically, FMEA is conducted by an interdisciplinary team ensuring comprehensive consideration of system interactions. FMEA provides a semi-quantitative method, supporting risk assessment by the calculation of a risk priority number (RPN), which is derived from the estimated probability of occurrence, significance, and the error's detectability. The implementation of FMEA at an early stage in the development process has been shown to result in a substantial and quantifiable reduction in the likelihood of potential errors \cite{kok2023new}.
Originally developed by the US military, FMEA has become a widely adopted tool across various domains including the maritime sector \cite{narayanagounder2009new, el2023overview}.

Another common method is the \textbf{Hazard and Operability study (HAZOP)} which provides a systematic approach to identify potential hazards in systems of all kinds \cite{international2001iec}. The method was developed in the 1970s in the chemical industry and is now employed in numerous domains.
HAZOP is qualitative method that employs a systematic brainstorming approach utilizing keywords to investigate deviations from specified behavior.
An interdisciplinary team examines the system under consideration from different perspectives to identify potential causes of errors, their consequences, and countermeasures.

A method that is designed to analyze causal chains leading to harm is the \textbf{Fault Tree Analysis (FTA)}. FTA is a deductive top-down hazard analysis method which aims at identifying and evaluating combinations of faults and failures that can lead to a predefined undesired event, commonly referred to as 'top level event'. Using Boolean logic and a hierarchical structure of logical gates (e.g. AND, OR), FTA systematically decomposes system-level failures into basic events at component level. This approach enables both qualitative understanding and quantitative risk assessment, including the calculation of failure probabilities. Initially developed in the aerospace and nuclear industries \cite{vesely1981fault}, FTA has been widely adopted in various sectors. It is particularly valued for its structured reasoning, and ability to support quantitative risk assessment.
FTA is especially effective when applied to hardware-dominated systems. However, extensions have been developed to  address human-related hazards and SOTIF, such as provided by Birch et al.~\cite{birch_human_2023} or Kramer et al.~\cite{kramer2020identification}.

Similar to FTA, \textbf{Event tree analysis (ETA)} employs a logic tree structure  to model potential progression of accidents capturing system responses and failure chains.
In contrast to FTA, however, ETA relies on inductive reasoning determining possible outcomes that may result from a specific initiating event \cite{ericson2005eta}. 
ETA supports both qualitative and quantitative risk evaluation.

\textbf{Bayesian Networks (BNs)}, also known as belief networks, provide another method to investigate causal chains. BNs are probabilistic graphical models that represent variables and their conditional dependencies using directed acyclic graphs \cite{pearl88, pearl2009}. In BNs, each node corresponds to a system variable, while the edges represent statistical or causal dependencies, quantified through conditional probability tables. BNs enable reasoning under uncertainty by relying on principles of probability theory, making them highly suitable for complex systems that are employed  in uncertain environments. BNs were originally developed for decision support and diagnostics and have been adapted for use in various safety-critical domains. Their ability to integrate both expert judgment and empirical data makes them particularly valuable for probabilistic risk assessment of dynamic and complex systems.

\textbf{System-Theoretic Process Analysis (STPA)} developed by Leveson and Thomas \cite{HandbookSTPA} is a relatively novel hazard analysis method grounded in system theory. 
It conceptualizes safety as a control problem emphasizing inadequate control actions within socio-technical systems rather than isolated component failures.
This perspective makes STPA especially well-suited for systems with complex interactions and software components. 
STPA employs a top-down approach consisting of four main steps: 
First, the goals of the analysis are defined in form of losses. 
In the second step, the system is modeled in form of a hierarchic control structure.
Based on this model, unsafe control actions are identified using a keyword-based technique. 
Finally, so called loss scenarios are derived containing causal factors that may lead to the unsafe control actions.
Originally developed for aerospace, STPA has since been adapted across various industries. 
In the maritime domain, STPA has gained relevance in the context of autonomous vessels \cite{Zhou2020, Li2023}.

A framework that has been developed by EMSA particularly for the hazard analysis of MASS is the \textbf{Risk-based Assessment Tool (RBAT)}. RBAT provides a structured methodology to compare automation and remote operations safety with conventional shipping \cite{emsa2024rbat}.
The methodology consists of five main parts and a total of 19 steps, encompassing the description of automation usage, hazard identification, mitigation analysis, to risk assessment and risk control. Central to RBAT is the modeling of vessel missions, control functions and the qualitative assessment of risks.
Risk levels for each scenarios are derived from a combination of worst-case outcome severity, the effectiveness of mitigation measures and the vessels exposure to enabling conditions.
Rather than focusing on the probability of systematic failures, the method integrates technical and operational aspects and emphasizes minimizing the consequences of functional failures.
A key feature of RBAT is the explicit integration of Remote Operation Centers (ROC) as supervisory unit within the safety analysis. The methodology enables the systematic identification of scenarios, in which the ROC is required to intervene, the information it must receive to perform this role, and the system architecture necessary to support interventions -- particularly in relation to mitigation strategies. Supervisory control agents located within the ROC operate in either an active or passive monitoring capacity, typically involving human operators. The effectiveness of mitigation measures attributed to the ROC is thus evaluated not only on the technical basis, but also with regard to human performance factors, such as operator response time or workload.

Another method  specifically developed for automated systems has been proposed by Kramer et al.~\cite{kramer2020identification,bode2019identifikation}. The \textbf{Automation Risk} method has initially been designed for automotive applications, but has also been transferred to the maritime domain \cite{vander2019approach, Hake25}. It aims to identify and evaluate hazardous scenarios, thereby supporting a scenario-based safety assessment. Conceptually, the method draws upon HAZOP and FTA, adapting and integrating elements of both to address the specific challenges that arise for highly automated systems. 

\subsection*{Human Factors}
Human factors risk analysis methods have become essential for understanding and mitigating the impact of human error in complex systems. These methods recognize that performance is shaped by a combination of individual capabilities, organizational culture, environmental influences, and system design. Traditional engineering risk assessment techniques often fall short in capturing these human and organizational dimensions, prompting the development of dedicated methodologies. Broadly speaking, human factors analysis approaches can be classified into predictive methods, which aim to anticipate potential errors during system design, and retrospective methods -- like HFAC \cite{wiegmann2003human} or HFACS-MA \cite{CHEN2013105} -- that analyze incidents after they occur. This paper focuses on predictive methods suitable for integration into early system development phases.

One such predictive methodology is the \textbf{Systematic Human Error Reduction and Prediction Approach (SHERPA)}, which provides a structured framework for anticipating human errors during task performance. Introduced by Embrey \cite{embrey1986sherpa}, SHERPA employs hierarchical task analysis to decompose complex operations into subtasks and applies error mode identification to foresee potential failure modes. Its strength lies in its proactive application during system design, helping to prevent errors by addressing both internal human factors and external, error-promoting conditions.

\textbf{THERP}, or the \textbf{Technique for Human Error Rate Prediction}, offers a quantitative means of evaluating human reliability, especially in high-risk settings like nuclear power. Developed by Swain and Guttmann \cite{swain1983technique}, this method integrates task analysis with human error probabilities and performance-shaping factors to deliver probabilistic risk estimates. While modeling human variability remains a challenge, THERP’s primary value is in supplying numerical data to broader system reliability assessments.

The \textbf{Functional Resonance Analysis Method (FRAM)} is presented by Hollnagel \cite{hollnagel2012fram} as a paradigm shift from traditional accident analysis methods toward understanding complex socio-technical systems. Resilience engineering has consistently argued that safety is more than the absence of failures, and FRAM builds on this foundation. FRAM is based on four principles: the equivalence of failures and successes, the central role of approximate adjustments, the reality of emergence, and functional resonance as a complement to causality. Unlike conventional methods that focus on what went wrong, FRAM is used to model the functions that are needed for everyday performance to succeed, and this model can then be used to explain specific events by showing how functions can be coupled. Over the past two decades, systemic-based risk assessment methods have garnered more attention, and FRAM is one of the most widely used systemic methods for risk assessment and accident analysis. The method represents Hollnagel's evolution from the more traditional CREAM approach \cite{hollnagel1998cognitive} toward understanding how normal performance variability can lead to both successful and unsuccessful outcomes in complex systems.

In the medical domain, human error analysis has evolved to include integrated techniques like the \textbf{Human Factors Failure Mode and Effects Analysis (HF-FMEA)}. Song et al.~\cite{SONG2020105918} demonstrate how combining traditional FMEA with human factors considerations enhances safety in medical device usage. Their approach allows for systematic identification of possible user-related errors, risk evaluation, and prioritization of preventive measures, adapting a well-established reliability tool to address human contributions more directly.

Another valuable technique, the \textbf{Success Likelihood Index Method (SLIM)}, is used to assess human reliability in specialized maritime operations such as pilot transfers. Aydin and colleagues \cite{AYDIN2022112830} show how SLIM, when integrated with the HFACS-PV framework, enables both qualitative and quantitative analysis of performance-shaping factors. This combination helps to clarify mechanisms behind human error while also estimating error probabilities, contributing to comprehensive maritime safety assessments.

In the petroleum sector, \textbf{Petro-HRA (Petroleum Human Reliability Assessment)} has been developed as an industry-specific method tailored to offshore operations. As outlined by Blackett et al.~\cite{blackett2022petro}, this approach enables both qualitative and quantitative evaluation of tasks affecting major accident risk. Its emphasis on post-initiating event scenarios, complex technical systems, and harsh operational environments underscores the limitations of generic HRA tools and the value of specialized adaptations.

The \textbf{Analysis of Pre-Accident Operator Actions (APOA)} offers another perspective by focusing on human actions occurring prior to accident events. Øie and Fernander \cite{apoa} introduce APOA as a structured method for tracing the sequence of decisions and actions that influence accident development. By examining the timing and context of human involvement, this approach enhances understanding of how specific actions may either exacerbate or mitigate incident outcomes, particularly in petroleum and maritime contexts.

The \textbf{CRIOP (Crisis Intervention and Operability Analysis)} framework is adopted by Hoem, Rødseth, and Johnsen \cite{hoem2021} as an interdisciplinary risk analysis method specifically applied to the design of remote control centers for maritime autonomous systems. The authors demonstrate how CRIOP can be effectively utilized to identify and analyze human factors risks in the emerging field of autonomous maritime operations, where traditional shipboard crew operations are replaced by shore-based remote monitoring and control. Their work shows how the CRIOP framework addresses the unique challenges of designing human-machine interfaces and operational procedures for remote maritime operations, considering both technical system capabilities and human operator competency requirements. The paper illustrates the framework's value in bridging the gap between human factors analysis and system design in advanced maritime technologies, providing a structured approach to ensure that remote control centers are designed with appropriate consideration of human performance limitations and requirements.

Taken together, these methodologies illustrate the expanding toolkit available for human factors analysis. Their diverse strategies -- from structured task analysis to probabilistic modeling and cognitive frameworks -- highlight the critical importance of anticipating human error during system design. By integrating these predictive methods early in the development process, industries can better manage safety risks and improve system resilience across a range of high-hazard domains.

%% file: sections/03_ROC.tex

Following the literature review of \autoref{sec:related_work}, we now move to our modeling activities.
In \autoref{subsec:ROC_introduction}, we introduce maritime ROCs as preparation for the conceptual description of the hazard database. As a illustrating use-case we describe berthing in a port in \autoref{subsec:use_case}. This leads to a generic MASS-ROC functional architecture described in  \autoref{subsec:functional_architecture}.

\subsection{Introduction to Maritime ROCs}
\label{subsec:ROC_introduction}

A ROC is a shore-based control center required for monitoring, controlling and supporting MASS. Depending on the level of automation, MASS can be highly automated or autonomous surface ships that can be remotely monitored and controlled by a ROC.
Although the ROC operators are physically separated from the ship, they perform important tasks during regular operations, in extreme situations, and when making critical decisions \cite{ISOTS23860}.
The main features of a ROC are:
\begin{itemize}
    \item[$\bullet$] there is a human operator (HO)
    \item[$\bullet$] it centralizes the monitoring and remote control of MASS
    \item[$\bullet$] its operating modes range from passive monitoring to active remote control of the ship
    \item[$\bullet$] it offers a high degree of automation on board with minimal human intervention if required,
    \item[$\bullet$] it has technical interfaces to navigation systems, sensors, communication, emergency management.
\end{itemize}
Importantly, the ISO/TS 23860 specifies four different operation modes for ROCs \cite{ISOTS23860}:
\begin{description}
\item[\textbf{Strategic Control}:] In this mode, the operator provides instructions to the entire fleet. This covers planning and organisational tasks. For example, a strategy for saving fuel under certain conditions may be communicated. 
\item[\textbf{Tactical Control}:] We are now moving to the individual MASS level. Tactical control is used to influence the decision making of an automation system. In contrast to the long-term approach of strategic control, tactical control takes a more short-term view. For example, decisions on routing or adjusting speeds in cooperation with the automated system (which directly controls the MASS).
\item[\textbf{Direct Control}:] Direct control means interacting directly with the functions of the MASS. This would directly override the decisions of an automation system. The operator literally controls the MASS. This includes all parameters and processes that can be manipulated and controlled. For example direct remote control of the MASS. It would be possible to take over the control directly in difficult situations, such as lock passages or at berthing places where the limits of the automation have been reached. In order to take over direct control from automation to operator, coordination must be carried out within the context of Human-Automation cooperation.
\item[\textbf{Monitoring}:] Monitoring includes the observation and evaluation of the MASS (the ship and the automation) and the environment or situation in which the MASS is located.
The aim is therefore to recognise deviations or anomalies in order to be able to react to them. Operators in the ROC monitor by receiving information about relevant processes via displays and control panels. Alarms also help them to quickly draw attention to a deviation or anomaly.
\end{description}
\begin{figure}[h!]
    \centering
    \includegraphics[width=0.6\columnwidth]{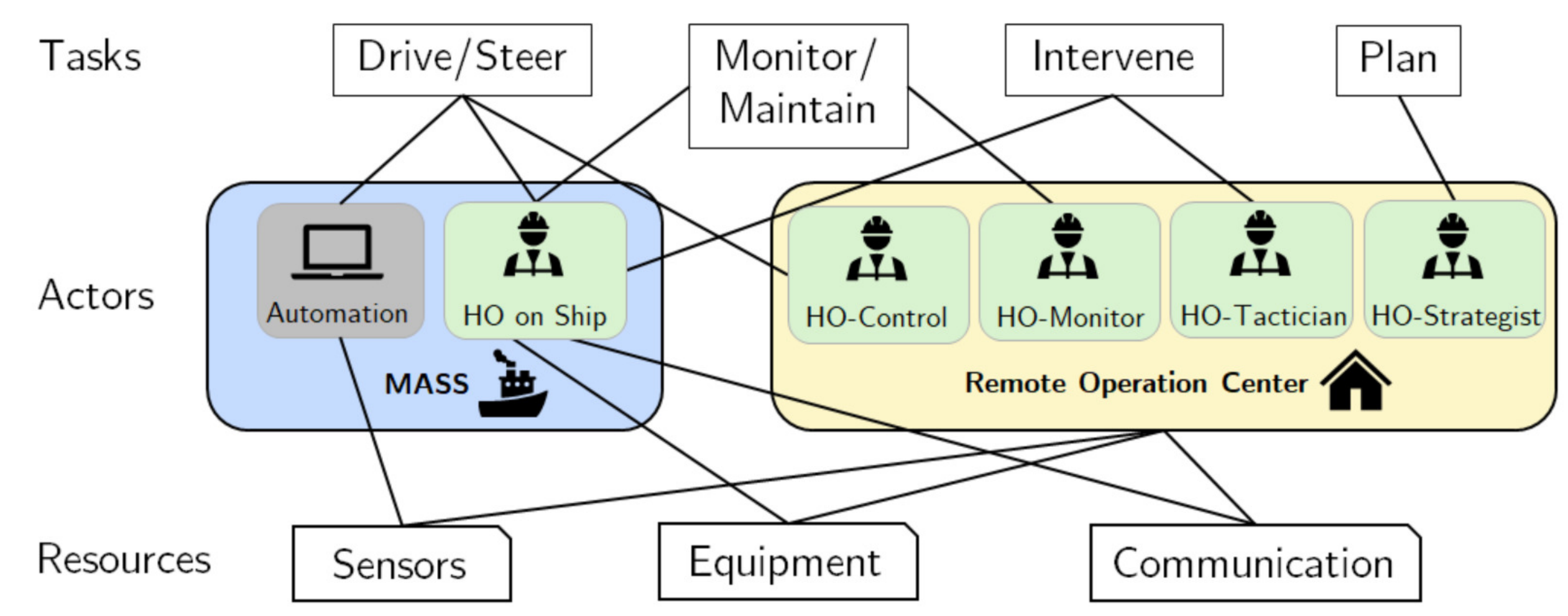}
    \caption{Tasks, Actors, and Resources derived from ISO/TS 23860.}
    \label{fig:actors_ROC}
\end{figure}
Figure \ref{fig:actors_ROC} shows the ROC from a human factors perspective. The different control modes are represented as potential actors. Although the crew on board and the automation of the MASS are located outside the ROC, they are important actors to the overall concept. Depending on the degree of automation, a crew on board may be optional.

\subsection{Use Case Example: Takeover Request during Port Entrance}
\label{subsec:use_case}

\begin{figure}[h!]
    \centering
    \includegraphics[width=0.6\columnwidth]{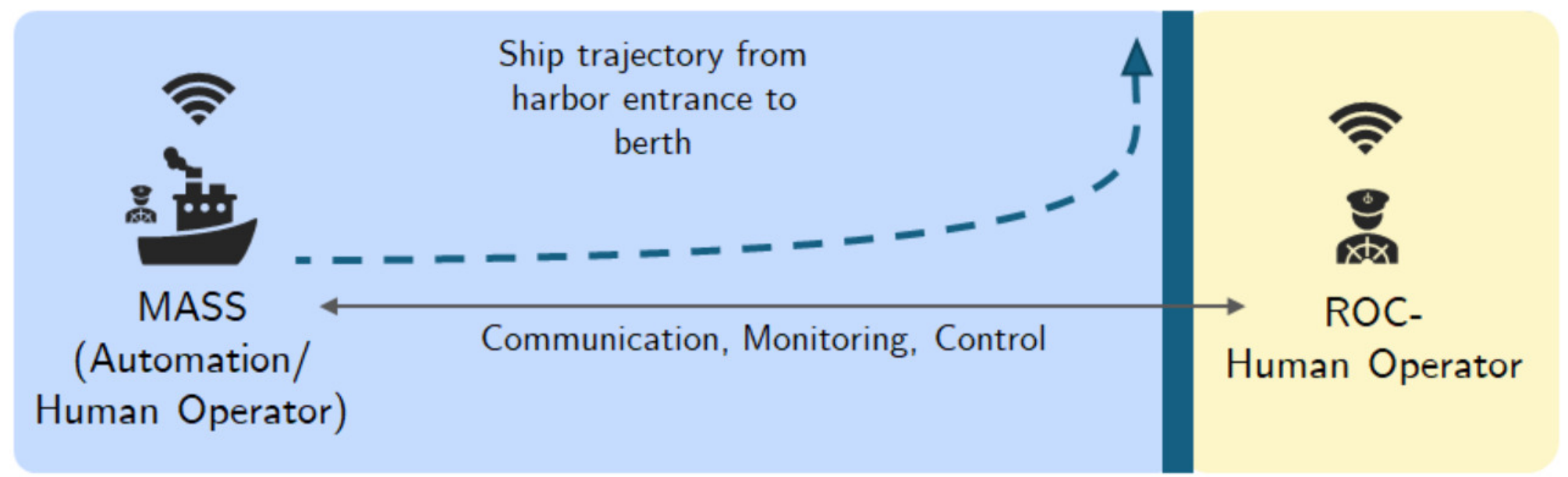}
    \caption{Use case: MASS maneuvering from harbor entrance to berth, adapted from Saager et al.~\cite{Saager2024}.}
    \label{fig:usecase_ROC}
\end{figure}

To substantiate our approach, we sketch a use case for which a MASS-ROC certification may be required. Consider a shore-based ROC controlling a MASS in a harbor where berthing is generally difficult due to its geometry. For our use case we assume that a MASS -- controlled by the automation -- 
approaches the harbor area and wants to dock at the berth. In addition, there is a crew on board that can take over some nautical or technical tasks if required. The ROC operator assists the ship in safely entering the port and during the berthing maneuver. Here, support is provided either tactically by intervening in the automation, or -- if necessary -- by directly controlling individual ship functions remotely.
Figure \ref {fig:usecase_ROC} depicts this use case which is considered a typical process occurring in daily ship operations.
The goal is to certify this MASS-ROC pair for this use case efficiently.
%

Next, we want to model a functional architecture for this use case. This forms the basis for a subsequent HARA and also for a potential database scheme. The results can therefore fill the hazard database with content for that use case. This, in turn, aids the certification process for other ROC-MASS pair in this use case.

\subsection{Functional Architecture}
\label{subsec:functional_architecture}

We start by modeling the functional architecture of a generic MASS-ROC pair at a high level of abstraction with a focus on the flow of information. This functional architecture, shown in Figure~\ref{fig:functional_architecture}, serves as a starting point for building a hazard database as it defines generic interfaces between the involved entities.
Note that the abstraction level needs to be detailed enough to enable the identification and analysis of hazards and abstract enough to keep HARA efforts manageable.
Inside the ROC itself, we model exactly one control station with three components:

\begin{figure}[h!]
    \centering
    \includegraphics[width=\columnwidth]{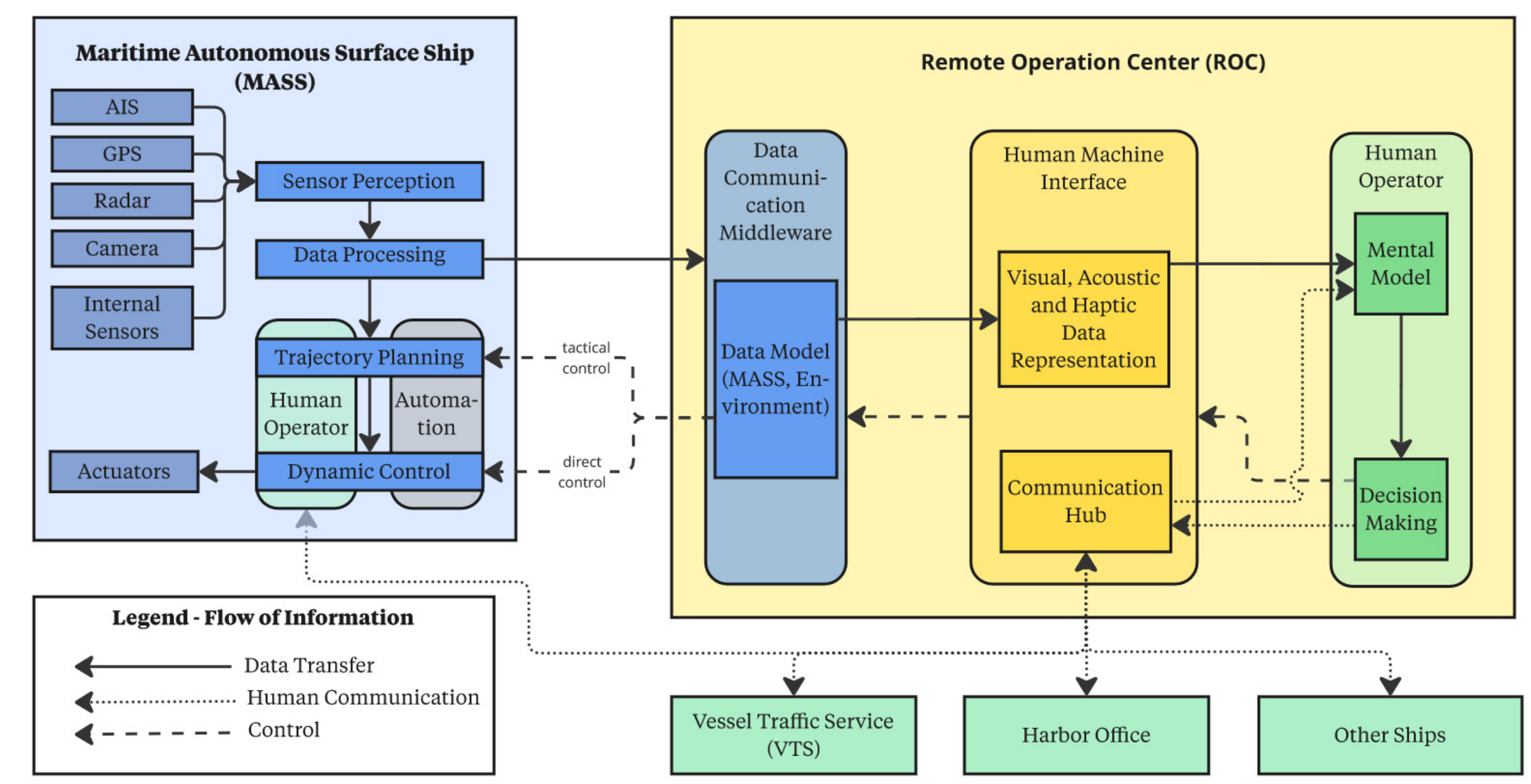}
    \caption{Functional architecture for a single control station inside a shore-based ROC and a generic MASS. The flow of information in encoded by three different types of arrows.}
    \label{fig:functional_architecture}
\end{figure}

\begin{description}
  \item[\textbf{Data Communication Middleware (DCM)}:] This component is responsible for managing the exchange of data between the ROC and the MASS. As indicated by the blue color coding, this component is present in the MASS as well. However, for brevity, we abstain from explicitly modeling it within the MASS here. Crucially, the DCM defines the interfaces between the ROC and the MASS. As input to the ROC we have the transfer of all relevant data from the MASS. This inncludes the MASS's data model and environment model. On the side of the ROC's output, the DCM facilitates remote control (direct or tactical) of the MASS by the ROC's human operator.
    \item[\textbf{Human Machine Interface (HMI)}:] The HMI bridges the gap between all inputs to the ROC and its outputs by interacting with the ROC's human operator. In particular, the HMI feeds the human operator's mental model with a visual, acoustic and haptic representation of data as well as human communication from external entities such as the MASS's human operator (if present), vessel traffic service (VTS), the harbor office, and other ships. His mental model infuses the decision making which directly leads to the human operator's output in form of either taking control of the MASS or communication with other humans - both of which are managed by the HMI's communication hub.
     \item[\textbf{Human Operator}:] The final component of the ROC is the human operator. All his actions within the ROC are through the HMI. His output is the HMI's input and vice versa as described before in section \ref{subsec:ROC_introduction}.
\end{description}

The functional architecture of Figure \ref{fig:functional_architecture} can now be used with various HARA methods. For example, it supports keyword-based approaches to hazard identification such as HAZOP -- which are also suggested by RBAT \cite[§~4.2]{emsa2024rbat}. 
Note that for a concrete ROC, the level of detail should be expanded locally when conductive for hazard identification and analysis. 
E.g., applying the keyword \emph{not provided} to \emph{visual, acoustic, and haptic representation of data}, a more detailed modeling of the HMI's data representation functionality becomes necessary.
If one wants to use STPA, a corresponding control loop can easily be derived from the architecture of Figure \ref{fig:functional_architecture}.
Moreover, methods for causal analysis such as FTA/ETA or causal Bayesian \cite{gansch2025causal} networks profit greatly from a functional architecture, because it supports the modeling of the system's internal dependencies. 


%% file: sections/04_HARA.tex

Similar to the database of criticality phenomena for automated driving systems suggested in previous work, cf.~Neurohr et al.\ \cite [§~4.2.1]{neurohr_2024_advances} and Babisch et al.\ \cite{babisch2023leveraging}, we propose to collect generic safety artifacts for maritime ROCs in a \emph{Hazard-DB}. This Hazard-DB can include suitable abstractions of
\begin{itemize}
    \item[$\bullet$] hazards and their potential sources,
    \item[$\bullet$] the corresponding causal relations \cite{koopmann2025grasping},
    \item[$\bullet$] risks and harms associated with these hazards, and
    \item[$\bullet$] strategies and mechanisms for risk mitigation.
\end{itemize}
When such safety-relevant artifacts have been identified and analyzed during a HARA for a concrete MASS-ROC pair, they can be integrated into the Hazard-DB, cf.~\autoref{fig:hazard_database}. This includes an appropriate abstraction step, as the goal is to reuse these artifacts for future HARAs.

\subsection{Categorization of Hazard Sources}
\label{subsec:categorization}

In order to gain an initial structure for HARA artifacts within a hazard database, we derive the following categorization of sources of hazard directly from the functional architecture of Figure \ref{fig:functional_architecture}:
\begin{description}
    \item[Data Communication:] This category includes technical hazards sources in the data communication between the ROC and the MASS on a technical level. Examples would be a disturbed and non-redundant communication channel, software failures in the DCM, but also include erroneous sensor data due to perception failures on the MASS. 
    \item[Human Machine Interface:] This category contains technical hazards sources originating in the HMI located in the ROC. Examples include the malfunctioning of displays, software failures related to the user interface, or alarm system failures.   
    \item[Human Operator:] This category consists of hazards originating from the behavior of the human operating the ROC. Included here are errors of commission, errors of omission, lack in skills (e.g.\ remote operating via a joystick, lacking knowledge of vessel), erroneous interpretation of data, and misuse of the HMI.
\end{description}

Additionally, we argue that the \textbf{Safety of the Intended Functionality} (SOTIF) is relevant for ROC safety and, as a category, complementary to the above. 
SOTIF covers hazards that are caused by so called functional insufficiencies, i.e. limitations of the technical capabilities and insufficiencies of the specification, including the inability to handle reasonably foreseeable misuse, as well as overall insufficiencies in the HMI design (e.g., too small fonts, too low contrast, cluttering, and information overload).
Depending on the type of remote operation, the human operator is responsible for monitoring, direct, strategic or tactical control. 
As the operator is not on the vessel, all decisions must be derived based on data acquired through some kind of sensor input.
Thus, similar to automated drivings systems governed by the ISO~21448 standard \cite{iso21448}, it is imperative to ensure that the sensor data and their subsequent processing deliver sufficient information to enable, in this case the human operator, to effectively execute their designated tasks. 

Viewing our categorization of hazard sources for maritime ROCs through the ISO~21448's lens, we spot several potential connections.
Both, the technical data communication as well as the HMI are considered to be within the scope of SOTIF.
Functional insufficiencies concerning the technical communication may include, for example, an inadequate handling of signal dead zones or inference caused by signal jamming.
For the HMI, functional insufficiencies may concern the information presented to the HO, e.g., an inadequate warning presented data are outdated.
Moreover, in the SOTIF context, the human operator's behavior is analyzed in terms of direct and indirect misuse.
We remark that all these considerations are applicable for maritime ROC safety.

\subsection{Concept for a Hazard Database}

\begin{figure}[h!]
    \centering
    \includegraphics[width=0.7\columnwidth]{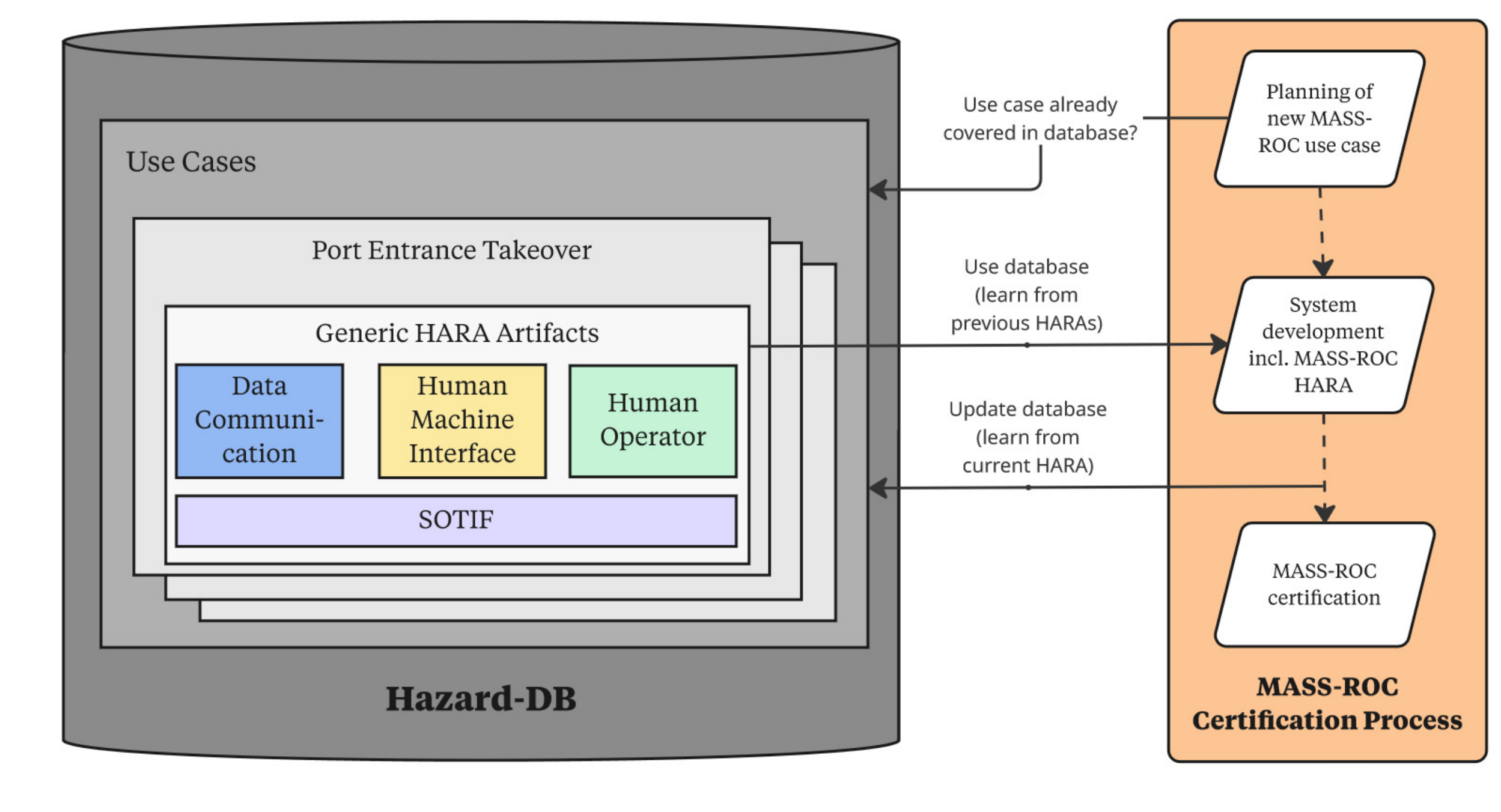}
    \caption{Concept for a hazard database supporting the certification process of a MASS-ROC pair.}
    \label{fig:hazard_database}
\end{figure}

Based on the categorization of hazard sources we provide a first sketch of how a hazard database could support the certification process of a MASS-ROC pair, cf.~Figure \ref{fig:hazard_database}.
The Hazard-DB contains use cases, exemplified by 'Port Entrance Takeover', which encompasses generic HARA artifacts rooted in the categories from the functional architecture of Figure~\ref{fig:functional_architecture}: Data Communication, Human Machine Interface, Human Operator, and SOTIF. The certification workflow demonstrates a cyclical process beginning with the planning of new MASS-ROC use cases, progressing through system development with an integrated HARA, and culminating in the MASS-ROC certification. 

The concept employs a bidirectional knowledge transfer mechanism: it leverages existing database knowledge from previous HARAs to inform current analyses while simultaneously updating the database with insights from ongoing HARA processes, thereby creating a continuous learning framework that enhances the accuracy and comprehensiveness of future risk assessments for MASS-ROC pairs. Once the Hazard-DB underwent the inital set up for a given use case, we can subsequently expect an efficiency increase regarding the future certification processes.

\subsection{Preliminary Suitability Analysis}
\label{subsec:suitability_analysis}

In order to grasp which of the HARA methods of \autoref{sec:related_work} have the potential to support the build-up of a hazard database, cf.~\autoref{fig:hazard_database}, we performed a preliminary, expert-based evaluation of their applicability to the three categories of hazard sources plus SOTIF. For each method, we evaluated whether it is applicable for technical hazard sources (i.e., data communication and human machine interface), human hazard sources, and whether SOTIF aspects are addressed appropriately.

Table \ref{subsec:suitability_analysis} shows the results of this preliminary suitability analysis.  A green check mark means that the method supports the category by design, a orange check mark means 
the method has been extended to this category, and a red cross encodes that the method has not been 
applied yet. Note that we did not evaluate to what degree the methods (or their extensions) cover these categories.

\begin{table}[h!]
    \caption{Overview which HARA methods address the introduced categories of hazard sources. A green check mark means "supported", an orange check mark means "could be extended to", and a red cross means "not supported".}
	\centering
	\footnotesize
	\renewcommand*{\arraystretch}{1.2}
	\begin{tabular}{l C{2.3cm} C{2.3cm} C{1.5cm} C{1.5cm}}
	    \toprule
	    & Data Communi\-cation & Human Machine Interface & Human Operator & SOTIF Aspects \\
	    \midrule
	   STPA \cite{HandbookSTPA} & \color{green}{\cmark} & \color{green}{\cmark} & {\color{orange}{\cmark}} \cite{france2017engineering}& {\color{orange}{\cmark}}\cite{iso21448} \\
	   FMEA \cite{SAEJ1739}& \color{green}{\cmark} & \color{green}{\cmark} & {\color{orange}{\cmark}} \cite{SONG2020105918} &  \color{red}{\xmark} \\
	   FTA \cite{vesely1981fault} & \color{green}{\cmark} & \color{green}{\cmark} &  {\color{orange}{\cmark}} \cite{birch_human_2023} & {\color{orange}{\cmark}} \cite{kramer2020identification} \\
	   ETA \cite{ericson2005eta} & \color{green}{\cmark} & \color{green}{\cmark} & {\color{orange}{\cmark}} \cite{birch_human_2023}  & \color{red}{\xmark}  \\
	   HAZOP \cite{international2001iec} & \color{green}{\cmark} & \color{green}{\cmark} & {\color{orange}{\cmark}} \cite{dunjo_hazard_2010} & {\color{orange}{\cmark}} \cite{kramer2020identification} \\
	   RBAT \cite{emsa2024rbat} & \color{green}{\cmark} & \color{green}{\cmark} &  \color{red}{\xmark}   & \color{green}{\cmark} \\
	   Bayesian Networks \cite{pearl88}& \color{green}{\cmark} & \color{green}{\cmark} &  {\color{orange}{\cmark}} \cite{trucco2008human} & {\color{orange}{\cmark}} \cite{gansch2025causal} \\
	   Automation Risks \cite{kramer2020identification} & \color{green}{\cmark} & \color{green}{\cmark} & \color{red}{\xmark} & \color{green}{\cmark} \\
	   FRAM \cite{hollnagel2012fram}& \color{green}{\cmark} & \color{green}{\cmark} &  \color{green}{\cmark} & \color{red}{\xmark}  \\
	   SHERPA \cite{embrey1986sherpa}& \color{red}{\xmark} & \color{red}{\xmark} & \color{green}{\cmark} & \color{red}{\xmark}\\
	   THERP \cite{swain1983technique}& \color{red}{\xmark} & \color{red}{\xmark} & \color{green}{\cmark} & \color{red}{\xmark} \\
	   SLIM \cite{AYDIN2022112830}& \color{red}{\xmark} & \color{red}{\xmark} & \color{green}{\cmark} & \color{red}{\xmark} \\
	   Petro-HRA \cite{blackett2022petro} & \color{red}{\xmark} & \color{red}{\xmark} &  \color{green}{\cmark} & \color{red}{\xmark} \\
	   APOA \cite{apoa} & \color{red}{\xmark} & \color{red}{\xmark} &  \color{green}{\cmark} & \color{red}{\xmark} \\
	   CRIOP \cite{hoem2021} & \color{red}{\xmark} & \color{red}{\xmark} &  \color{green}{\cmark} & \color{green}{\cmark} \\
	    \bottomrule
	\end{tabular}
	\label{tab:analysis_methods}
\end{table}

Summarizing \autoref{subsec:suitability_analysis}, we see an expected divide between methods that natively cover technical hazard and those that were designed for human factors. Therefore, to generate results that densely fill the Hazard-DB, we want to either
\begin{itemize}
    \item[(i)] choose a universal method with adequate extensions to all categories such as STPA, or
    \item[(ii)] combine a technical method with a human factors method, e.g., RBAT and FRAM.
\end{itemize}
Finally, we want to emphasize the integration of SOTIF aspects for all three categories. Neglecting functional insufficiencies or foreseeable misuse, e.g., in the HMI design, can easily lead to accidents during operations.

%% file: sections/10_Conclusion.tex

In this paper, we laid first steps towards building a hazard database for certification of shore-based ROCs. Based on a generic MASS-ROC functional architecture we derived three different categories of hazard sources while identifying SOTIF as a relevant, complementary category. Moreover, we performed a preliminary suitability analysis of HARA methods which may cover these categories.

Regarding future work, conducting a HARA for a concrete shore-based ROC by combining adequate techniques will enable the initial build-up of safety artifacts for the envisioned Hazard-DB.

%% file: Literature.bib
@inproceedings{kramer2020identification,
	author={Kramer, Birte and Neurohr, Christian and Büker, Matthias and Böde, Eckard and Fränzle, Martin and Damm, Werner},
	editor={Zeller, Marc and Höfig, Kai},
	title={{Identification and Quantification of Hazardous Scenarios for Automated Driving}},
	booktitle={{Model-Based Safety and Assessment}},
	year={2020},
	publisher= {Springer International Publishing},
	address={Cham},
	pages={163--178},
	isbn={978-3-030-58920-2},
	doi={10.1007/978-3-030-58920-2_11}
}

@ARTICLE{koopmann2025grasping,
  author={Koopmann, Tjark and Putze, Lina and Westhofen, Lukas and Gansch, Roman and Adee, Ahmad and Neurohr, Christian},
  journal={IEEE Access}, 
  title={{Grasping Causality for the Explanation of Criticality for Automated Driving}}, 
  year={2025},
  volume={13},
  number={},
  pages={54739-54756},
  doi={10.1109/ACCESS.2025.3555177}}

@article{babisch2023leveraging,
author = {Babisch, Stefan and Neurohr, Christian and Westhofen, Lukas and Schoenawa, Stefan and Liers, Henrik},
title = {{Leveraging the GIDAS Database for the Criticality Analysis of Automated Driving Systems}},
journal = {Journal of Advanced Transportation},
volume = {2023},
number = {1},
pages = {1349269},
doi = {https://doi.org/10.1155/2023/1349269},
year = {2023}
}

@InProceedings{gansch2025causal,
author="Gansch, Roman
and Putze, Lina
and Koopmann, Tjark
and Reich, Jan
and Neurohr, Christian",
title={{Causal Bayesian Networks for Data-Driven Safety Analysis of Complex Systems}},
booktitle="Model-Based Safety and Assessment",
year="2026",
publisher="Springer Nature Switzerland",
address="Cham",
pages="222--237"
}

@techreport{bode2019identifikation,
  title={{Identifikation und Quantifizierung von Automationsrisiken f{\"u}r hochautomatisierte Fahrfunktionen}},
  author={B{\"o}de, Eckard and B{\"u}ker, Matthias and Damm, Werner and Fr{\"a}nzle, Martin and Kramer, Birte and Neurohr, Christian and Vander Maelen, Sebastian},
  year={2019},
  institution={Insitute for Information Technology (OFFIS e.V.)},
  publisher = {PEGASUS Project},
  doi = {10.5281/zenodo.14141022}
}

@misc{neurohr_2024_advances,
  author       = {Neurohr, Christian and
                  Westhofen, Lukas and
                  Butz, Martin and
                  Bollmann, Martin Herbert and
                  Putze, Lina and
                  Koopmann, Tjark and
                  Gansch, Roman and
                  Knoop, Michael and
                  Rasch, Armin and
                  Cojocaru, Bogdan and
                  Daube, Johannes},
  title        = {{Advances on the Criticality Analysis for Automated
                   Driving Systems}},
  month        = feb,
  year         = 2024,
  publisher    = {Zenodo},
  doi          = {10.5281/zenodo.10815308}
}

@mastersthesis{gali2024remote,
  title={{Remote Operation Centers for Autonomous Ships}},
  author={Gal{\'\i} Debouche, Carla and others},
  type={{B.Sc.} thesis},
  year={2024},
  school={Universitat Polit{\`e}cnica de Catalunya}
}

@article{bratic2019review,
  title={{A Review of Autonomous and Remotely Controlled Ships in Maritime Sector}},
  author={Brati{\'c}, Karlo and Pavi{\'c}, Ivan and Vuk{\v{s}}a, Sr{\dj}an and Stazi{\'c}, Ladislav},
  journal={Transactions on Maritime Science},
  volume={8},
  number={02},
  pages={253--265},
  year={2019},
  publisher={Sveu{\v{c}}ili{\v{s}}te u Splitu, Pomorski fakultet},
  doi={10.7225/toms.v08.n02.011}
}

@article{riedmaier2020survey,
  title={{Survey on Scenario-Based Safety Assessment of Automated Vehicles}},
  author={Riedmaier, Stefan and Ponn, Thomas and Ludwig, Dieter and Schick, Bernhard and Diermeyer, Frank},
  journal={IEEE access},
  volume={8},
  pages={87456--87477},
  year={2020},
  publisher={IEEE}
}

@article{dnv2018autonomous,
  title={Autonomous and remotely operated ships},
  author={DNV},
  journal={Class Guideline DNV-CG-264},
  year={2024}
}

@article{dnv2022competence,
  title={Competence of remote control centre operators},
  author={DNV},
  journal={Standard DNV-ST-0324},
  year={2022}
}

@article{dnv2021certification,
  title={Certification scheme for remote control centre operators},
  author={DNV},
  journal={Recommended Practice DNV-RP-0323},
  year={2021}
}

@article{DNVRUPT6CH12,
  title={{Part 6 Additional class notations, Chapter 12 Autonomy and remote operation}},
  author={DNV},
  journal={Rules for Classification},
  year={2024}
}

@INPROCEEDINGS{brinkmann2017learning,
  author={Brinkmann, Marius and Böde, Eckard and Lamm, Arne and Maelen, Sebastian Vander and Hahn, Axel},
  booktitle={OCEANS 2017 - Aberdeen}, 
  title={{Learning from automotive: Testing maritime assistance systems up to autonomous vessels}}, 
  year={2017},
  volume={},
  number={},
  pages={1-8},
  doi={10.1109/OCEANSE.2017.8084951}}

@INPROCEEDINGS{vander2019approach,
  author={Vander Maelen, Sebastian and Büker, Matthias and Kramer, Birte and Böde, Eckard and Gerwinn, Sebastian and Hake, Georg and Hahn, Axel},
  booktitle={2019 4th International Conference on System Reliability and Safety (ICSRS)}, 
  title={{An Approach for Safety Assessment of Highly Automated Systems Applied to a Maritime Traffic Alert and Collision Avoidance System}}, 
  year={2019},
  volume={},
  number={},
  pages={494-503},
  doi={10.1109/ICSRS48664.2019.8987712}}

@techreport{emsa2024rbat,
  title={{RBAT - Method Description}},
  author={Kvinnesland, Kenneth and Hoem, Asa Snilstveit and {{\O}}ie, Sondre  and Pederson, Remi Brensdal},
  year={2024},
  institution={European Maritime Safety Agency (EMSA)},
  publisher = {EMSA, Lisbon}
}

@article{wylie2024safety,
doi = {10.1088/1742-6596/2867/1/012045},
year = {2024},
month = {oct},
publisher = {IOP Publishing},
volume = {2867},
number = {1},
pages = {012045},
author = {Wylie, M and Rajabally, E},
title = {Safety Assurance of Maritime Autonomous Surface Ships},
journal = {Journal of Physics: Conference Series}
}

@article{wiegmann2003human,
  title={A human error approach to aviation accident analysis: The human factors analysis and classification system},
  author={Wiegmann, Douglas A and Shappell, Scott A},
  journal={Aviation, Space, and Environmental Medicine},
  volume={74},
  number={11},
  pages={1006--1016},
  year={2003},
  publisher={Aerospace Medical Association},
  notes={file:A Human Error Approach to Aviation Accident Analysis.pdf}
}

@article{embrey1986sherpa,
  title={{SHERPA}: A systematic human error reduction and prediction approach},
  author={Embrey, David},
  journal={Proceedings of the International Topical Meeting on Advances in Human Factors in Nuclear Power Systems},
  pages={184--193},
  year={1986}, 
  notes={file:SHERPA1986.PDF}
}

@techreport{swain1983technique,
  title={Handbook of Human Reliability Analysis with Emphasis on Nuclear Power Plant Applications},
  author={Swain, Alan David and Guttmann, H.E.},
  year={1983},
  institution={Sandia National Labs., Albuquerque, NM (USA)},
  notes={file:ML071210299.pdf}
}

@book{hollnagel1998cognitive,
  title={Cognitive reliability and error analysis method (CREAM)},
  author={Hollnagel, Erik},
  year={1998},
  publisher={Elsevier}
}

@book{hollnagel2012fram,
  title={FRAM: The Functional Resonance Analysis Method: Modelling Complex Socio-Technical Systems},
  author={Hollnagel, Erik},
  year={2012},
  publisher={Ashgate Publishing},
  isbn={978-1-4094-4551-7},
  address={Farnham, UK}, 
  doi={https://doi.org/10.1201/9781315255071}, 
  notes={file:FRAM_ The Functional Resonance Analysis Method _ Modelling.pdf}
}

@article{birch_human_2023,
	title = {Human {Reliability} {Analysis} using a {Human} {Factors} {Hazard} {Model}},
	volume = {58},
	url = {https://jsystemsafety.com/index.php/jss/article/view/251},
	doi = {10.56094/jss.v58i2.251},
	number = {2},
	journal = {Journal of System Safety},
	author = {Birch, Dustin and Miller, Erika and Bradley, Thomas},
	month = jun,
	year = {2023},
	pages = {7--29},
}

@INPROCEEDINGS{neurohr2020fundamental,
  author={Neurohr, Christian and Westhofen, Lukas and Henning, Tabea and de Graaff, Thies and Möhlmann, Eike and Böde, Eckard},
  booktitle={2020 IEEE Intelligent Vehicles Symposium (IV)}, 
  title={{Fundamental Considerations around Scenario-Based Testing for Automated Driving}}, 
  year={2020},
  volume={},
  number={},
  pages={121-127},
  doi={10.1109/IV47402.2020.9304823}}

@INPROCEEDINGS{putze2023validation,
  author={Putze, Lina and Westhofen, Lukas and Koopmann, Tjark and Böde, Eckard and Neurohr, Christian},
  booktitle={2023 IEEE Intelligent Vehicles Symposium (IV)}, 
  title={{On Quantification for SOTIF Validation of Automated Driving Systems}}, 
  year={2023},
  volume={},
  number={},
  pages={1-8},
  doi={10.1109/IV55152.2023.10186795}}

@article{SONG2020105918,
title = {Human factors risk assessment: An integrated method for improving safety in clinical use of medical devices},
journal = {Applied Soft Computing},
volume = {86},
pages = {105918},
year = {2020},
issn = {1568-4946},
doi = {https://doi.org/10.1016/j.asoc.2019.105918},
url = {https://www.sciencedirect.com/science/article/pii/S1568494619306994},
author = {Wenyan Song and Jing Li and Hao Li and Xinguo Ming},
notes = {file:1-s2.0-S1568494619306994-main.pdf}
}

@article{CHEN2013105,
title = {{A Human and Organisational Factors analysis method for marine casualties using HFACS-Maritime Accidents}},
journal = {Safety Science},
volume = {60},
pages = {105-114},
year = {2013},
issn = {0925-7535},
doi = {https://doi.org/10.1016/j.ssci.2013.06.009},
author = {Shih-Tzung Chen and Alan Wall and Philip Davies and Zaili Yang and Jin Wang and Yu-Hsin Chou}
}

@article{AYDIN2022112830,
title = {Assessment of human error contribution to maritime pilot transfer operation under {HFACS-PV} and {SLIM} approach},
author = {Muhammet Aydin and \"{O}zkan U\u{g}urlu and Muhammet Boran},
journal = {Ocean Engineering},
volume = {266},
pages = {112830},
year = {2022},
issn = {0029-8018},
doi = {https://doi.org/10.1016/j.oceaneng.2022.112830},
url = {https://www.sciencedirect.com/science/article/pii/S0029801822021138},
notes={file:1-s2.0-S0029801822021138-main.pdf}
}

@article{akyuz2017application,
title = {Application of fuzzy logic to fault tree and event tree analysis of the risk for cargo liquefaction on board ship},
journal = {Applied Ocean Research},
volume = {101},
pages = {102238},
year = {2020},
issn = {0141-1187},
doi = {https://doi.org/10.1016/j.apor.2020.102238},
url = {https://www.sciencedirect.com/science/article/pii/S0141118719310296},
author = {Emre Akyuz and Ozcan Arslan and Osman Turan}, 
notes={file:1-s2.0-S0141118719310296-main.pdf}
}

@article{trucco2008human,
  title={A Bayesian Belief Network modelling of organisational factors in risk analysis: A case study in maritime transportation},
  author={Trucco, Paolo and Cagno, Enrico and Ruggeri, Fabrizio and Grande, Oreste},
  journal={Reliability Engineering \& System Safety},
  volume={93},
  number={6},
  pages={845--856},
  year={2008},
  publisher={Elsevier}, 
  notes={file:1-s2.0-S0951832007001214-main.pdf}
}

@techreport{blackett2022petro,
author = {Blackett, Claire and Farbrot, Jan Erik and \O{}ie, Sondre and Fernander, Marius},
year = {2022},
pages = {1--99},
isbn={978-82-7017-937-4},
title = {{The Petro-HRA Guideline Rev.1 Vol. 1}}, 
institution = {IFE - Institute for Energy Technology}
}

@book{apoa, 
author = {\O{}ie, Sondre and Fernander, Marius},
title = {{Analysis of Pre-Accident Operator Actions (APOA)}},
month={August},
year={2023},
pages={67},
publisher={DNV},
country={Norway},
isbn={978-82-515-0325-9}, 
notes={file:APOA_Equinor_Energy_AS.pdf}
}

@article{Zhou2020,
title = {Towards applicability evaluation of hazard analysis methods for autonomous ships},
journal = {Ocean Engineering},
volume = {214},
pages = {107773},
year = {2020},
issn = {0029-8018},
doi = {https://doi.org/10.1016/j.oceaneng.2020.107773},
url = {https://www.sciencedirect.com/science/article/pii/S0029801820307502},
author = {Xiang-Yu Zhou and Zheng-Jiang Liu and Feng-Wu Wang and Zhao-Lin Wu and Ren-Da Cui},
}

@article{Li2023,
title = {Risk and reliability analysis for maritime autonomous surface ship: A bibliometric review of literature from 2015 to 2022},
journal = {Accident Analysis \& Prevention},
volume = {187},
pages = {107090},
year = {2023},
issn = {0001-4575},
doi = {https://doi.org/10.1016/j.aap.2023.107090},
url = {https://www.sciencedirect.com/science/article/pii/S0001457523001379},
author = {Zhihong Li and Di Zhang and Bing Han and Chengpeng Wan},
}

@InProceedings{Hoem2021,
author="Hoem, {\AA}sa S.
and R{\o}dseth, {\O}rnulf J.
and Johnsen, Stig Ole",
editor="Arezes, Pedro M.
and Boring, Ronald L.",
title="{Adopting the CRIOP Framework as an Interdisciplinary Risk Analysis Method in the Design of Remote Control Centre for Maritime Autonomous Systems}",
booktitle="Advances in Safety Management and Human Performance",
year="2021",
publisher="Springer International Publishing",
address="Cham",
pages="219--227",
isbn="978-3-030-80288-2"
}

@article{Thieme2018,
title = {{Assessing ship risk model applicability to Marine Autonomous Surface Ships}},
journal = {Ocean Engineering},
volume = {165},
pages = {140-154},
year = {2018},
issn = {0029-8018},
doi = {https://doi.org/10.1016/j.oceaneng.2018.07.040},
url = {https://www.sciencedirect.com/science/article/pii/S0029801818313210},
author = {Christoph Alexander Thieme and Ingrid Bouwer Utne and Stein Haugen},
}

@article{Wrobel2016,
	author = {Wrobel, Krzysztof and Krata, Przemyslaw and Montewka, Jakub and Hinz, Tomasz},
	title = {Towards the Development of a Risk Model for Unmanned Vessels Design and Operations},
	journal = {TransNav, the International Journal on Marine Navigation and Safety of Sea Transportation},
	volume = {10},
	number = {2},
	pages = {267-274},
	year = {2016},
	url = {./Article_Towards_the_Development_of_a_Risk_Wrobel,38,648.html},
	doi = {10.12716/1001.10.02.09},
	issn = {2083-6473},
	publisher = {Gdynia Maritime University, Faculty of Navigation},
}

@article{Wrobel2018a,
title = {System-theoretic approach to safety of remotely-controlled merchant vessel},
journal = {Ocean Engineering},
volume = {152},
pages = {334-345},
year = {2018},
issn = {0029-8018},
doi = {https://doi.org/10.1016/j.oceaneng.2018.01.020},
url = {https://www.sciencedirect.com/science/article/pii/S0029801818300209},
author = {Krzysztof Wróbel and Jakub Montewka and Pentti Kujala},
}

@article{Wrobel2018b,
title = {{Towards the development of a system-theoretic model for safety assessment of autonomous merchant vessels}},
journal = {Reliability Engineering \& System Safety},
volume = {178},
pages = {209-224},
year = {2018},
issn = {0951-8320},
doi = {https://doi.org/10.1016/j.ress.2018.05.019},
author = {Krzysztof Wróbel and Jakub Montewka and Pentti Kujala}
}

@proceedings{Utne2017,
    author = {Utne, Ingrid Bouwer and Sørensen, Asgeir J. and Schjølberg, Ingrid},
    title = {{Risk Management of Autonomous Marine Systems and Operations}},
    volume = {Volume 3B: Structures, Safety and Reliability},
    series = {International Conference on Offshore Mechanics and Arctic Engineering},
    year = {2017},
    month = {06},
    organization = {ASME - American Society of Mechanical Engineers},
    doi = {10.1115/OMAE2017-61645}
}

@article{Banda2019,
title = {A systemic hazard analysis and management process for the concept design phase of an autonomous vessel},
journal = {Reliability Engineering \& System Safety},
volume = {191},
pages = {106584},
year = {2019},
issn = {0951-8320},
doi = {https://doi.org/10.1016/j.ress.2019.106584},
url = {https://www.sciencedirect.com/science/article/pii/S0951832017314151},
author = {Osiris A. {Valdez Banda} and Sirpa Kannos and Floris Goerlandt and Pieter H.A.J.M. {van Gelder} and Martin Bergström and Pentti Kujala},
}

@MISC{iso26262,
	author = {{International Organization for Standardization}},
	publisher = {{ISO, Geneva, Switzerland}},
	title = {{ISO 26262: Road vehicles -- Functional safety}},
	type = {{Standard}},
	year = {{2018}}
}

@MISC{iso21448,
	author = {{International Organization for Standardization}},
	publisher = {{ISO, Geneva, Switzerland}},
	title = {{ISO 21448: Road vehicles -- Safety of the intended functionality}},
	type = {{Standard}},
	year = {{2022}}
}

@MISC{ISOTS23860,
	author = {{International Organization for Standardization}},
	publisher = {{ISO, Geneva, Switzerland}},
	title = {{ISO/TS 23860: Ships and marine technology — Vocabulary related to autonomous ship systems}},
	type = {{Standard}},
	year = {{2022}}}

@MISC{SAEJ1739,
	author = {{SAE International}},
	publisher = {{United States, Department of Defense}},
	title = {{SAE Standard J1739\_202101 Potential Failure Mode and Effects Analysis (FMEA) Including Design FMEA, Supplemental FMEA-MSR, and Process FMEA}},
	type = {{Standard}},
	year = {{2021}},
	doi={10.4271/J1739_202101}
}

@article{Saager2024,
    title = {{Towards Modelling Cooperation in Future Maritime Remote-Control Center}},
    journal = {Design for Equality and Justice. INTERACT 2023. Lecture Notes in Computer Science},
    year = {2024},
    issn = {},
    doi = {https://doi.org/10.1007/978-3-031-61688-4_29},
    author = {Saager, Marcel and Harre, Marie-Christin and Hahn, Axel},
}

@techreport{emsa2023CMOROC,
  title={{ CMOROC Identification of Competences for MASS Operators in Remote
Operation Centre}},
  author={Jung, Thomas and Harre, Marie-Christin and Rousselle, Noelle and Lüedtke, Andreas and Saager, Marcel},
  year={2023},
  institution={European Maritime Safety Agency (EMSA)},
  publisher = {EMSA, Lisbon}
}

@techreport{vesely1981fault,
  title={Fault tree handbook},
  author={Vesely, William E and Goldberg, Francine F and Roberts, Norman H and Haasl, David F},
  year={1981},
  institution={Nuclear Regulatory Commission Washington DC}
}

@article{narayanagounder2009new,
  title={A new approach for prioritization of failure modes in design FMEA using ANOVA},
  author={Narayanagounder, Sellappan and Gurusami, Karuppusami},
  journal={World Academy of Science, Engineering and Technology},
  volume={49},
  number={524-31},
  year={2009}
}

@article{el2023overview,
  title={Overview of failure mode and effects analysis (FMEA): a patient safety tool},
  author={El-Awady, Shaymaa MM},
  journal={Global Journal on Quality and Safety in Healthcare},
  volume={6},
  number={1},
  pages={24--26},
  year={2023},
  publisher={Innovative Healthcare Institute}
}

@article{kok2023new,
  title={New generation FMEA method in automotive industry: an application},
  author={K{\"o}k, Nesimi and Y{\i}ld{\i}z, Mehmet Selami},
  journal={Journal of Turkish Operations Management},
  volume={7},
  number={1},
  pages={1630--1643},
  year={2023},
  publisher={Ankara Yildirim Beyazit University}
}

@article{murata89,
  author={Murata, T.},
  journal={{Proceedings of the IEEE}}, 
  title={Petri nets: Properties, analysis and applications}, 
  year={1989},
  volume={77},
  number={4},
  pages={541-580},
  doi={10.1109/5.24143}
  }

@book{pearl88,
    author = {Pearl, Judea},
    title = {Probabilistic Reasoning in Intelligent Systems: Networks of Plausible Inference},
    year = {1988},
    isbn = {1558604790},
    publisher = {Morgan Kaufmann Publishers Inc.},
    address = {San Francisco, CA, USA},
}

@article{international2001iec,
  title={{IEC 61882: Hazard and operability studies (HAZOP studies)--Application guide}},
  author={International Electrotechnical Commission and others},
  journal={International Electrotechnical Commission, Geneva, Switzerland},
  year={2001}
}

@article{dunjo_hazard_2010,
	title = {Hazard and operability ({HAZOP}) analysis. {A} literature review},
	volume = {173},
	issn = {0304-3894},
	doi = {10.1016/j.jhazmat.2009.08.076},
	journal = {Journal of Hazardous Materials},
	author = {Dunjó, Jordi and Fthenakis, Vasilis and Vílchez, Juan A. and Arnaldos, Josep},
	year = {2010},
	pages = {19--32}
}

@incollection{ericson2005eta,
author = {Ericson, Clifton A. II},
publisher = {John Wiley \& Sons, Ltd},
isbn = {9780471739425},
title = {{Event Tree Analysis}},
booktitle = {{Hazard Analysis Techniques for System Safety}},
chapter = {12},
pages = {223-234},
doi = {https://doi.org/10.1002/0471739421.ch12},
year = {2005}
}

@book{HandbookSTPA,
  title = {{STPA} {Handbook}},
  author = {Leveson, Nancy G. and Thomas, John P.},
  year = {2018},
  publisher = {MIT - Massachusetts Institute of Technology},
  urldate = {2024-07-11},
url = {http://psas.scripts.mit.edu/home/get_file.php?name=STPA_Handbook.pdf}
}

@phdthesis{france2017engineering,
  title={{Engineering for humans: A new extension to STPA}},
  author={France, Megan Elizabeth},
  year={2017},
  school={Massachusetts Institute of Technology}
}

@book{pearl2009,
  place={Cambridge},
  edition={2},
  title={Causality}, 
  publisher={Cambridge University Press}, 
  author={Pearl, Judea}, 
  year={2009}
}

@misc{IMO2019,
  author = {{International Maritime Organization (IMO)}},
  title = {{Interim Guidelines for MASS Trials}},
  year = {2019},
  publisher = {International Maritime Organization, London},
  url = {https://wwwcdn.imo.org/localresources/en/MediaCentre/HotTopics/Documents/MSC.1-Circ.1604%20-%20Interim%20Guidelines%20For%20Mass%20Trials%20%28Secretariat%29.pdf},
  urldate = {2025-07-30}
}

@misc{IMO2025,
  author = {{International Maritime Organization (IMO)}},
  title = {{Maritime Safety Committee - 110th session (MSC 110), 18-27 June 2025}},
  year = {2025},
  url = {https://www.imo.org/en/mediacentre/meetingsummaries/pages/msc-110th-session.aspx},
  note = {Accessed on 2025-07-30},
  urldate = {2025-07-30}
}

@article{Hake25,
    title={{Safety Assessment of Maritime Autonomous Surface Ships: A Scenario-Based Approach}},
    author={Georg Hake and Jan Steffen Becker and Anna Austel and Lina Putze and Nina Wetzig},
    journal={Journal of Physics: Conference Series},
    year={in press},
}
